\documentclass[lettersize,journal]{IEEEtran}
\usepackage{amsmath,amsfonts}
\usepackage{algorithmic}
\usepackage{algorithm}
\usepackage{array}
\usepackage{textcomp}
\usepackage{stfloats}
\usepackage{url}
\usepackage{verbatim}
\usepackage{graphicx}
\usepackage{cite}
\usepackage{caption}
\usepackage{subcaption}
\begin{document}

\title{Target-Enclosed Least-Squares Seismic Imaging}

\author{Aydin Shoja, Joost van der Neut, Kees Wapenaar
\thanks{This work was supported by the European Union's Horizon 2020 Research and Innovation Program: European Research Council under Grant 742703. (\textit{Corresponding author: Aydin Shoja.})}%
\thanks{Aydin Shoja and Kees Wapenaar are with the Department of Geoscience and Engineering, Delft University of Technology, 2600 Delft, The Netherlands. (e-mail: s.m.a.shoja@tudelft.nl; c.p.a.wapenaar@tudelft.nl).}
\thanks{Joost van der Neut is with the Department of Imaging Physics, Delft University of Technology, 2600 Delft, The Netherlands (e-mail: j.r.vanderneut@tudelft.nl).}}

\markboth{IEEE transactions on geosciences and remote sensing}%
{Shoja \MakeLowercase{\textit{et al.}}: Target-enclosed least-squares seismic imaging}

\IEEEpubid{1558-0644 © 2023 IEEE. Personal use is permitted, but republication/redistribution requires IEEE permission. See https://www.ieee.org/publications/rights/index.html}

\maketitle

\begin{abstract}
Least-Squares Reverse-Time Migration (LSRTM) is a method that seismologists utilize to compute a high-resolution subsurface image. Nevertheless, LSRTM is a computationally demanding problem. One way to reduce the computational costs of the LSRTM is to choose a small region of interest and compute the image of that region. However, finding representations that account for the wavefields entering the target region from the surrounding boundaries is necessary. This paper confines the region of interest between two boundaries above and below this region. The acoustic reciprocity theorem is employed to derive representations for the wavefields at the upper and lower boundaries of the target region. With the help of these representations, a target-enclosed LSRTM algorithm is developed to compute a high-resolution image of the region of interest. Moreover, the possibility of using virtual receivers created by Marchenko redatuming is investigated.
\end{abstract}


\section{Introduction}\label{intro}
\IEEEPARstart{W}{avefield} migration is the art of computing the medium reflectivity from the recorded wavefield passing through the medium. Many migration algorithms are available, such as Kirchhoff migration \cite{Claerbout}, one-way wave equation migration \cite{Mulder}, and Reverse-Time Migration (RTM) \cite{BaysalRTM,Zhou,Zhang}. RTM is one of the common migration algorithms. RTM is commonly implemented by applying the adjoint of the Born operator to the recorded data \cite{SeisInv}. However, the inverse of the operator is needed for the true image. Consequently, the migration result suffers from amplitude and resolution issues \cite{Dutta,Tang,Liu}. One way to address this problem is to solve the migration problem with a least-squares solution called Least-Squares Reverse-Time Migration (LSRTM). The least-squares solution is usually applied as an iterative optimization problem.

However, LSRTM is computationally expensive \cite{Dai,Tang,Herrmann,Farshad}. To reduce the computation cost of the LSRTM, one can reduce the computation domain by confining the model to a Region of Interest (ROI) by recording or computing the wavefields at the boundary of this region. The process of migrating for a ROI is called target-oriented migration \cite{Valenciano,Haffinger,Willemsen,Yuan,Zhao,Guo2,Ravasi2}. However, deploying receivers on the boundaries of ROI usually is not possible due to physical obstacles. 

Due to the aforementioned accessibility issue, the typical approach for target-oriented migration is to opt for redatuming algorithms and only consider the upper horizontal boundary of the ROI or target \cite{Guo1,Guo2,Wapenaar6,Vargas,Schuster,Ravasi1,Luiken,Shoja3}. In cases where the ROI is enclosed between two boundaries, i.e. when wavefields are entering the ROI from the underburden through the lower boundary \cite{Wapenaar2016,Weglein}, the shortcoming of only considering the upper boundary is that any wavefield entering the ROI from the medium below the lower boundary of the ROI is unaccounted for, hindering the convergence of the inversion process. Moreover, including the lower boundary in the algorithm can add transmission information to the inversion. However, including the lower boundary in the inversion process has rarely been studied directly. For instance, Cui et al. \cite{Cui} derive a representation with a reciprocity theorem and Marchenko redatuming to include surrounding boundaries in target-oriented full waveform inversion (FWI), Diekmann et al. \cite{Diekman1} use a Marchenko retrieved Green's function of the ROI and insert it in the Lippmann-Schwinger integral to create a linear inversion process, and van der Neut et al. \cite{Joost1} design a target-enclosed imaging algorithm with the help of a reciprocity theorem. Of the above-mentioned papers, only \cite{Joost1} directly studies the consequences of including the lower boundary in the imaging process, and the others only implicitly imply the effects of it.

This paper, which is an extension of \cite{Joost1}, studies the contribution of the lower boundary by introducing a target-enclosed LSRTM algorithm. To derive this algorithm, we start by explaining the LSRTM briefly. Next, we derive a target-enclosed representation for Green's functions on the upper and lower boundaries of the ROI by using the reciprocity theorem. Then, we connect this target-enclosed representation to LSRTM to complete our algorithm. After deriving the required equations, we test our algorithm with numerical examples. First, we use physical receivers at the boundaries of the ROI to check the results in the ideal situation. Then, we briefly introduce Marchenko redatuming to explore the possibility of using virtual receivers in our algorithm. For both physical and virtual receivers cases we use both a homogeneous background and a smooth background velocity as the velocity model for migration. Finally, we finish the paper by discussing the results and providing a conclusion.

\section{\label{sec:2} Theory}
To develop the theory, we start with a brief discussion of LSRTM. Next, a representation of the target-enclosed Green's function is given with the help of the reciprocity theorem. Finally, to derive our target-enclosed LSRTM formulation, we combine LSRTM with the target-enclosed representations. \newpage In the entire theory section, we are in the frequency-space domain, and for simplicity, we drop the dependency on angular frequency ($\omega$).

\subsection{Least-Squares Reverse-Time Migration}
We start the explanation of LSRTM by investigating the Born integral for the scattered wavefield by a scattering potential \cite{born_wolf_1999,SeisInv,Peter}. Here we follow the convention of \cite{Peter}:
  \begin{equation}
  P^{scat}(\textbf{x}')=\int_{\mathcal{V}}\gamma_0^2(\textbf{x})G_{0}(\textbf{x}',\textbf{x})\chi^c(\textbf{x})P^{inc}(\textbf{x})d\textbf{x}.
  \label{scattering}
  \end{equation}
In this equation, $\textbf{x}'$ is the observation location, $\textbf{x}$ is a location inside the computation volume ($\mathcal{V}$), $P^{scat}(\textbf{x}')$ is the scattered pressure field at the observation point, $P^{inc}(\textbf{x})$ is the incident pressure field at the computation point and $G_{0}(\textbf{x}',\textbf{x})$ is the background Green's function between $\textbf{x}$ and $\textbf{x}'$. Moreover, $\gamma_0(\textbf{x})=\frac{-i\omega}{c_0(\textbf{x})}$ and $\chi^c(\textbf{x})=1-\frac{c^2_0(\textbf{x})}{(c^{scat}(\textbf{x}))^2}$ is the propagation velocity perturbation, where $c_0(\textbf{x})$ and $c^{scat}(\textbf{x})$ are the background and scatterer's velocity, respectively.

It is possible to rewrite this equation in matrix form:
  \begin{equation}
    \textbf{P}_{pred}^{scat}(\delta \textbf{m})=\textbf{L} \delta \textbf{m}.
    \label{scatt_mat}
  \end{equation}
Here, $\textbf{L}$ is matrix form of the integral operator of equation \ref{scattering} and $\delta\textbf{m}$ is a vector, containing the perturbation $\chi^c(\textbf{x})$.

To obtain an estimation of $\delta\textbf{m}$, we can apply the adjoint of $\textbf{L}$ to the observed data:
  \begin{equation}
    \delta \textbf{m}^{img}=\textbf{L}^\dagger \textbf{P}_{obs}^{scat},
    \label{imaging}
  \end{equation}
  where $\dagger$ denotes complex conjugate transposition.
  
 We can go one step further to obtain a high-resolution estimation of the image by minimizing the following objective function \cite{Marquardt}:
  \begin{equation}
    J(\delta \textbf{m})=\frac{1}{2}\|\textbf{P}_{pred}^{scat}(\delta \textbf{m})-\textbf{P}_{obs}^{scat}\|^2_2.
    \label{costfunc}
  \end{equation}
  Different optimization algorithms, such as conjugate gradient, can minimize this function. This optimization problem is known as Least-Squares Reverse-Time Migration (LSRTM).
 
  \subsection{Target-enclosed representations}\label{TE}
  LSRTM is a computationally expensive algorithm. In order to reduce its computational burden, seismologists usually opt for a target-oriented algorithm, limiting the medium to a smaller region. As we mentioned in section \ref{intro}, target-oriented algorithms usually redatum the data to the upper boundary of the target and ignore any information coming from the lower boundary.
  
\begin{figure}
\centering
\begin{subfigure}{0.5\textwidth}
\centering
\includegraphics[width=0.6\textwidth]{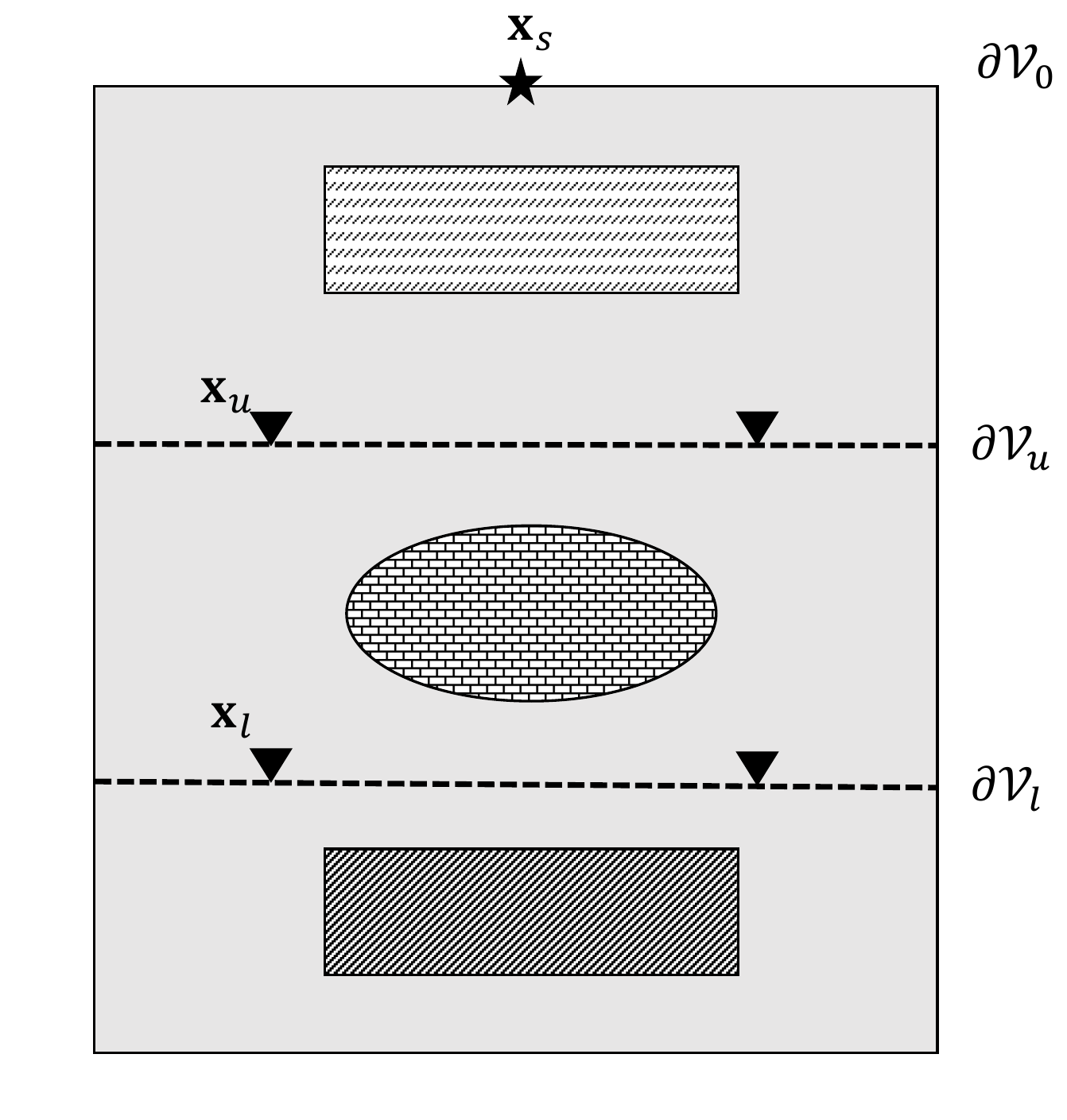}
\caption{}
\label{stateA}
\end{subfigure}
\hfill
\begin{subfigure}{0.5\textwidth}
\centering
\includegraphics[width=0.6\textwidth]{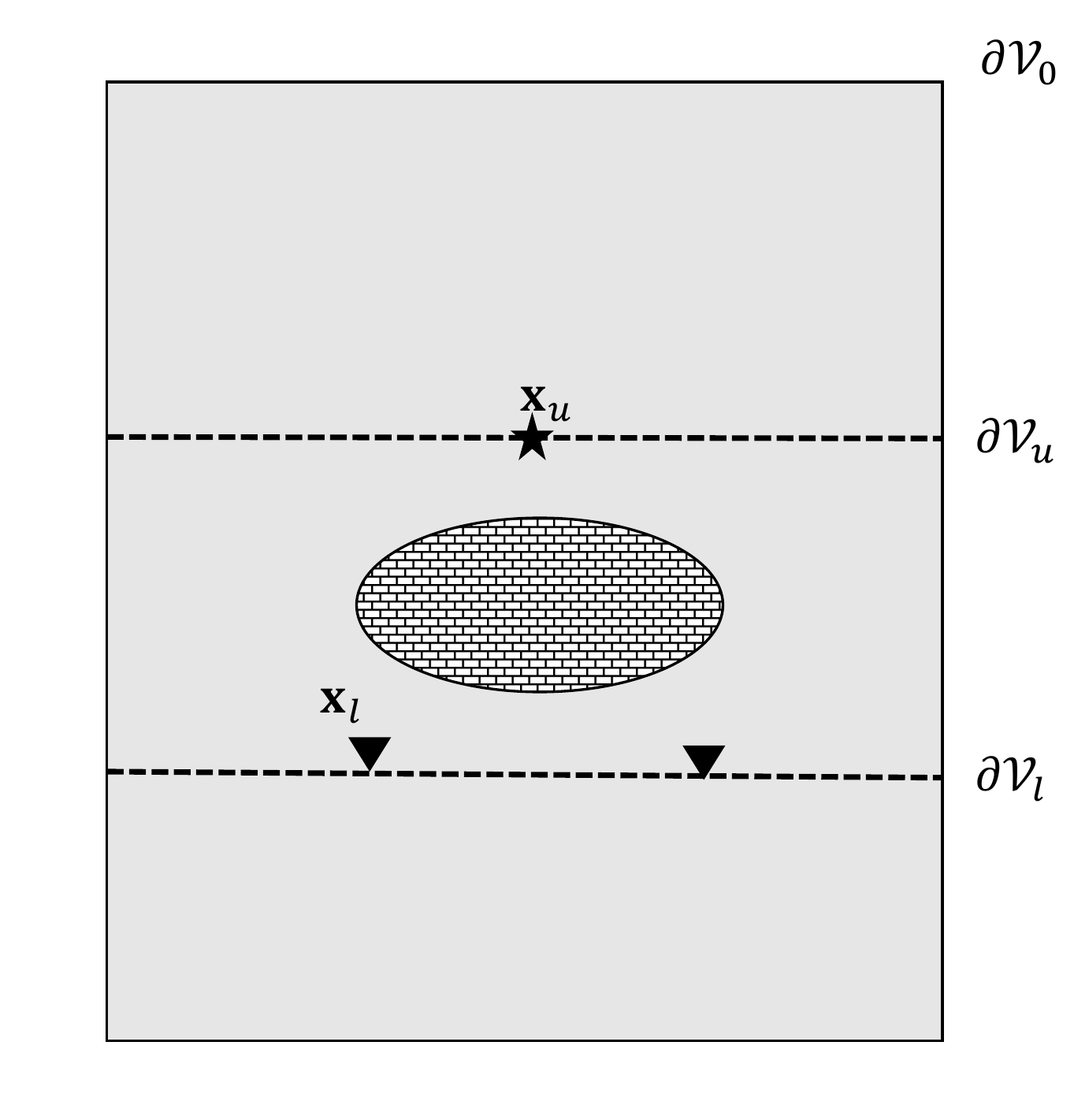}
\caption{}
\label{stateB}
\end{subfigure}
\caption{ (a) State A and (b) state B. Black stars depict source locations, and black reversed triangles depict receiver locations.}
\label{States}
\end{figure}

  In this section, the idea is to find a representation that can account for a heterogeneous medium above the upper and below the lower boundary of the target area. To derive this representation, we follow \cite{Joost1}. The starting point of the derivation is the acoustic reciprocity theorem of the convolution type \cite{Wapenaar6}, which connects the wavefields of two different states via
  \begin{multline}
  \int_{\mathfrak{d}\mathcal{V}_u}\rho^{-1}(p^+_A(\partial_3p^-_B)+p^-_A(\partial_3p^+_B))d\textbf{x}= \\
  \int_{\mathfrak{d}\mathcal{V}_l}\rho^{-1}(p^+_A(\partial_3p^-_B)+p^-_A(\partial_3p^+_B))d\textbf{x}. 
  \label{reciprocity}
  \end{multline}
  Here, we consider a volume $\mathcal{V}$, which is limited by two infinite horizontal surfaces. States A and B (Fig. \ref{States}) are defined in two different media which are identical inside the volume $\mathcal{V}$ with boundaries denoted by $\mathfrak{d}\mathcal{V}_u$ (upper) and $\mathfrak{d}\mathcal{V}_l$ (lower), and arbitrary outside of this volume. In addition, $p^+$ and $p^-$ are decomposed wavefields on the boundaries where $+$ means downgoing and $-$ means upgoing, and $\partial_3$ is the partial derivative in direction $x_3$ (downward).
  
  To continue the derivation we define Green's function $G(\textbf{x},\textbf{x}_s)$ as the solution of the following Helmholtz equation:
  \begin{equation}
    \nabla^2 G + k(\textbf{x})^2 G = -\rho \delta(\textbf{x}-\textbf{x}_s),
     \label{waveeq}
  \end{equation}
  where $k(\textbf{x})=\frac{\omega}{c(\textbf{x})}$ is the wavenumber, and $c$ is the propagation velocity.

    A representation for Green's function at the upper boundary of the target ($\mathfrak{d}\mathcal{V}_u$) can be derived by defining state A in the actual medium and state B in a medium identical to medium A inside the volume $\mathcal{V}$ (target) and homogeneous outside of it. We denote the Green's functions of state A with $G(\textbf{x},\textbf{x}_s)$ and Green's function of state B with $G_{tar}(\textbf{x},\textbf{x}_u)$. In state A we define $p_A^\pm=G^\pm(\textbf{x},\textbf{x}_s)$, where $\textbf{x}_s$ is a location at Earth's surface $\mathfrak{d}\mathcal{V}_0$. For state B we have $p_B^\pm=G_{tar}^\pm(\textbf{x},\textbf{x}_u)$. Here, $\textbf{x}_u$ is located at the upper boundary $\mathfrak{d}\mathcal{V}_u$ and $G_{tar}^\pm(\textbf{x},\textbf{x}_u)$ is stimulated by an impulsive point source at $\textbf{x}_u$. Since the half-space above the target is homogeneous in state B, the vertical derivative of the $G_{tar}^+(\textbf{x},\textbf{x}_u)$ at $\mathfrak{d}\mathcal{V}_u$ is
  \begin{equation}
    \lim_{x_{u,3}\to x_{u,3^+}}\partial_3 G_{tar}^+(\textbf{x},\textbf{x}_u)=-\frac{\rho}{2}\delta(\textbf{x}_{\rm H}-\textbf{x}_{u,{\rm H}}).
  \label{dG+}
  \end{equation}
  Here, ${x_{u,3}\to x_{u,3^+}}$ means the limit from below the boundary, where $\textbf{x}_H$ and $\textbf{x}_{u,H}$ denote the horizontal coordinates of $\textbf{x}$ and $\textbf{x}_u$, respectively. In addition, since in state B the medium below the target is reflection-free, $G^-_{tar}(\textbf{x},\textbf{x}_u)$ and its derivative disappear at the lower boundary $\mathfrak{d}\mathcal{V}_l$.
  By substituting all of the ingredients into equation \ref{reciprocity}, the following can be reached:
  \begin{multline}
  G^-(\textbf{x}_u,\textbf{x}_s)=\int_{\mathfrak{d}\mathcal{V}_u}G^+(\textbf{x},\textbf{x}_s)\frac{2\partial_3}{\rho(\textbf{x})}G^-_{tar}(\textbf{x},\textbf{x}_u)d\textbf{x} \\
  +\int_{\mathfrak{d}\mathcal{V}_l}G^-(\textbf{x},\textbf{x}_s)\frac{2\partial_3}{-\rho(\textbf{x})}G^+_{tar}(\textbf{x},\textbf{x}_u)d\textbf{x}.
  \label{representation}
  \end{multline}
This equation is the base for our target-enclosed LSRTM derivation. The first integral on the right-hand side of Equation \ref{representation} accounts for anything entering the medium from the upper boundary, and the second integral accounts for anything that comes from the lower boundary.
    
  \subsection{Target-enclosed LSRTM}
\begin{figure}
\centering
\begin{subfigure}{0.5\textwidth}
\centering
\includegraphics[width=0.6\textwidth]{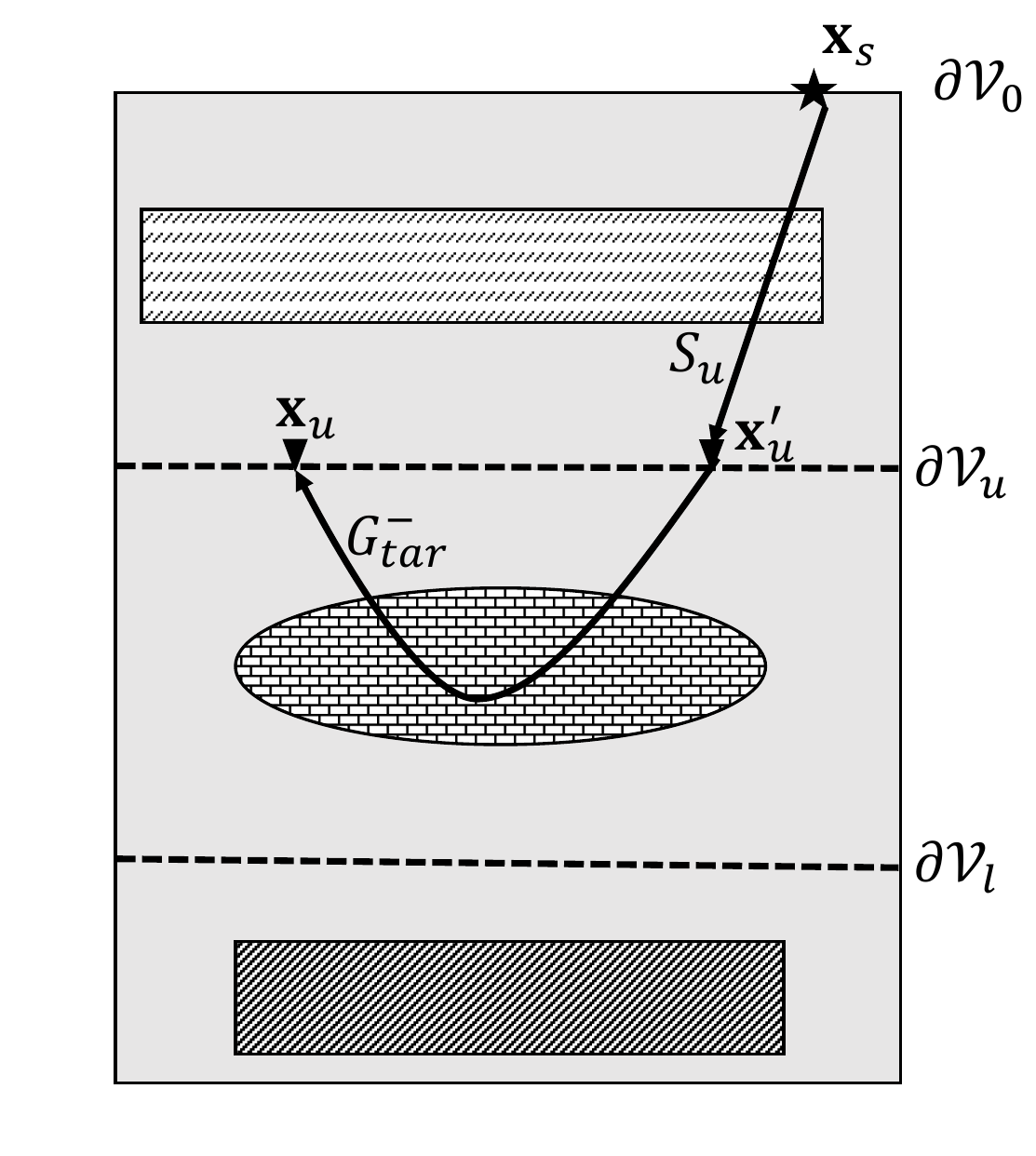}
\caption{}
\label{integralup}
\end{subfigure}
\hfill
\begin{subfigure}{0.5\textwidth}
\centering
\includegraphics[width=0.6\textwidth]{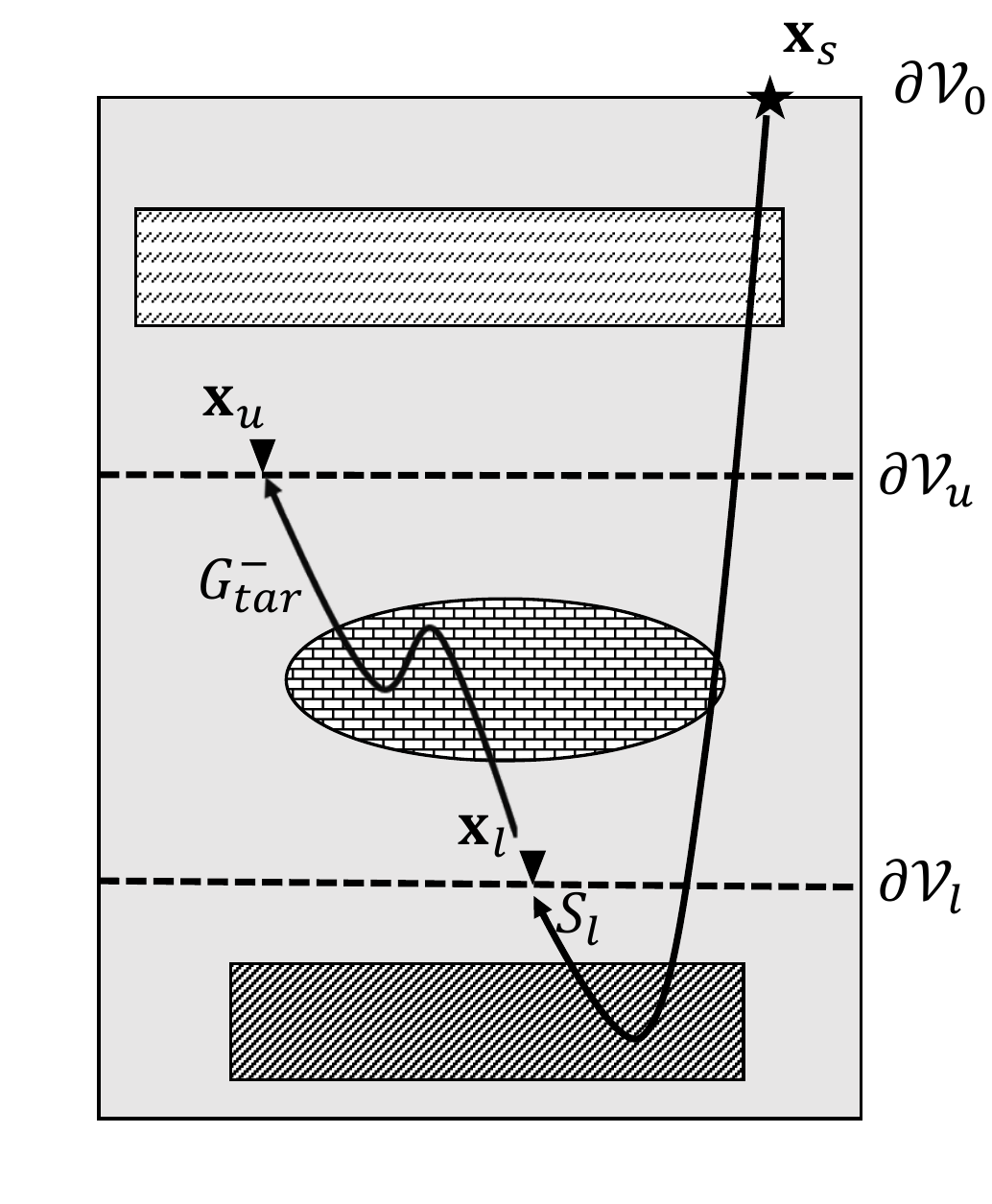}
\caption{}
\label{integrallow}
\end{subfigure}
\caption{(a) The first integral on the right-hand side of Equation \ref{representation_alt}, and (b) the second integral on the right-hand side of Equation \ref{representation_alt}.}
\label{integrals}
\end{figure}

To merge this representation with LSRTM, we use
  \begin{equation}
  \begin{split}
  \int_{\mathfrak{d}\mathcal{V}}p^\pm_A(\textbf{x})\partial_3p^\mp_B(\textbf{x}) d\textbf{x} = -\int_{\mathfrak{d}\mathcal{V}}p^\mp_B(\textbf{x})\partial_3p^\pm_A(\textbf{x}) d\textbf{x} 
  \end{split}
  \label{transpose}
  \end{equation}
 \cite{Keesbook}, and we use Green's function reciprocity:
  \begin{equation}
    G^-_{tar}(\textbf{x}'_u,\textbf{x}_u) = G^-_{tar}(\textbf{x}_u,\textbf{x}'_u)
  \label{Gupreciprocity}
  \end{equation}
  \begin{equation}
    G^+_{tar}(\textbf{x}_l,\textbf{x}_u) = G^-_{tar}(\textbf{x}_u,\textbf{x}_l),
  \label{Glowreciprocity}
  \end{equation}
 where $\textbf{x}'_u$ is an element of $\mathfrak{d}\mathcal{V}_u$ and $\textbf{x}_l$ is an element of $\mathfrak{d}\mathcal{V}_l$. We rewrite Equation \ref{representation} as follows:
  \begin{multline}
    G^-(\textbf{x}_u,\textbf{x}_s)W(\omega)=\int_{\mathfrak{d}\mathcal{V}_u}G^-_{tar}(\textbf{x}_u,\textbf{x}'_u)S_u(\textbf{x}'_u,\textbf{x}_s)d\textbf{x}'_{u} \\
    +\int_{\mathfrak{d}\mathcal{V}_l}G^-_{tar}(\textbf{x}_u,\textbf{x}_l)S_l(\textbf{x}_l,\textbf{x}_s)d\textbf{x}_{l},
  \label{representation_alt}
  \end{multline}
where $W(\omega)$ is the source signature, 
  \begin{equation}
  S_u(\textbf{x}'_u,\textbf{x}_s)=\frac{2\partial'_{3,u}}{-\rho(\textbf{x}'_u)}G^+(\textbf{x}'_u,\textbf{x}_s)W(\omega)
  \label{Source_udf}
  \end{equation}
is the dipole source from the upper boundary of the target, which accounts for reflections from above the upper boundary of the target. Further
  \begin{equation}
    S_l(\textbf{x}_l,\textbf{x}_s)=\frac{2\partial_{3,l}}{\rho(\textbf{x}_l)}G^-(\textbf{x}_l,\textbf{x}_s)W(\omega)
   \label{Source_ldf}
   \end{equation}
is the dipole source term from the lower boundary of the target, which accounts for the reflections generated below the target. Fig. \ref{integrals} represents the right-hand side of Equation \ref{representation_alt}. 

As said above, the second integral on the right-hand side of equation \ref{representation_alt} is the contribution of the medium below the target to the data. This integral can be split into two terms: 1) the arrival from the lower boundary to the upper one in the background, and 2) the forward scatterings inside the target region. In mathematical terms:
  \begin{multline}
  \int_{\mathfrak{d}\mathcal{V}_l}G^-_{tar}(\textbf{x}_u,\textbf{x}_l)S_l(\textbf{x}_l,\textbf{x}_s)d\textbf{x}_{l}= \\
  \int_{\mathfrak{d}\mathcal{V}_l}G^{-}_{0,tar}(\textbf{x}_u,\textbf{x}_l)S_l(\textbf{x}_l,\textbf{x}_s)d\textbf{x}_{l} \\
  +\int_{\mathfrak{d}\mathcal{V}_l}G^{-,scat}_{tar}(\textbf{x}_u,\textbf{x}_l)S_l(\textbf{x}_l,\textbf{x}_s)d\textbf{x}_{l}.
    \label{lower}
  \end{multline}
Here, $G^{-}_{0,tar}(\textbf{x}_u,\textbf{x}_l)$ is the Green's function in the background model of the target, and $G^{-,scat}_{tar}(\textbf{x}_u,\textbf{x}_l)$ is the Green's function that contains the scattered events. By substituting equation \ref{lower} in equation \ref{representation_alt} and taking the background contribution to the left-hand side we end up with:
  \begin{multline}
  G^-(\textbf{x}_u,\textbf{x}_s)W(\omega)-\int_{\mathfrak{d}\mathcal{V}_l}G^{-}_{0,tar}(\textbf{x}_u,\textbf{x}_l)S_l(\textbf{x}_l,\textbf{x}_s)d\textbf{x}_{l}= \\
  \int_{\mathfrak{d}\mathcal{V}_u}G^{-,scat}_{tar}(\textbf{x}_u,\textbf{x}'_u)S_u(\textbf{x}'_u,\textbf{x}_s)d\textbf{x}'_{u} \\
  +\int_{\mathfrak{d}\mathcal{V}_l}G^{-,scat}_{tar}(\textbf{x}_u,\textbf{x}_l)S_l(\textbf{x}_l,\textbf{x}_s)d\textbf{x}_{l}.
  \label{modifired_representation_df}
  \end{multline}
Importantly, the Green's function $G^-_{tar}(\textbf{x}_u,\textbf{x}'_u)$ in the first integral on the right-hand side of equation \ref{representation_alt} is the scattered Green's function inside the target, so we rename it to $G^{-,scat}_{tar}(\textbf{x}_u,\textbf{x}'_u)$. 
  
To obtain our target-enclosed LSRTM algorithm, we take three steps: First, we assign
  \begin{multline}
   P_{obs}^{scat,TE}(\textbf{x}_u,\textbf{x}_s)= \\
  G^-(\textbf{x}_u,\textbf{x}_s)W(\omega)-\int_{\mathfrak{d}\mathcal{V}_l}G^{-}_{0,tar}(\textbf{x}_u,\textbf{x}_l)S_l(\textbf{x}_l,\textbf{x}_s)d\textbf{x}_{l},
    \label{PobsTE}
  \end{multline}
where "TE" stands for "Target-Enclosed". Second we compute the scattered Green's functions ($G^{-,scat}_{tar}(\textbf{x}_u,\textbf{x}'_u)$ and $G^{-,scat}_{tar}(\textbf{x}_u,\textbf{x}_l)$) with equation \ref{scattering}
  \begin{multline}
    G^{-,scat}_{tar}(\textbf{x}_u,\textbf{x}'_u) = \\
  \int_{\mathcal{V}}\gamma_0^2(\textbf{x})G_{0,tar}(\textbf{x}_u,\textbf{x})\chi^c(\textbf{x})G_{0,tar}(\textbf{x},\textbf{x}'_u)d\textbf{x},
  \end{multline}
and
  \begin{multline}
    G^{-,scat}_{tar}(\textbf{x}_u,\textbf{x}_l) = \\
    \int_{\mathcal{V}}\gamma_0^2(\textbf{x})G_{0,tar}(\textbf{x}_u,\textbf{x})\chi^c(\textbf{x})G_{0,tar}(\textbf{x},\textbf{x}_l)d\textbf{x},
  \end{multline}
  where $\textbf{x}$ is a location inside the target volume. Using this in the right-hand side of Equation \ref{modifired_representation_df}, we obtain
  \begin{multline}
  \int_{\mathfrak{d}\mathcal{V}_u}G^{-,scat}_{tar}(\textbf{x}_u,\textbf{x}'_u)S_u(\textbf{x}'_u,\textbf{x}_s)d\textbf{x}'_{u} \\
  +\int_{\mathfrak{d}\mathcal{V}_l}G^{-,scat}_{tar}(\textbf{x}_u,\textbf{x}_l)S_l(\textbf{x}_l,\textbf{x}_s)d\textbf{x}_{l}= \\
  \int_{\mathcal{V}}\gamma_0^2(\textbf{x})G_{0,tar}(\textbf{x}_u,\textbf{x})\chi^c(\textbf{x})P_{inc,TE}(\textbf{x},\textbf{x}_s)d\textbf{x},
  \end{multline}
 where
  \begin{multline}
  P^{inc,TE}(\textbf{x},\textbf{x}_s)= \\
  \int_{\mathfrak{d}\mathcal{V}_u}G_{0,tar}(\textbf{x},\textbf{x}'_u)S_u(\textbf{x}'_u,\textbf{x}_s)d\textbf{x}'_{u} \\
  +\int_{\mathfrak{d}\mathcal{V}_l}G_{0,tar}(\textbf{x},\textbf{x}_l)S_l(\textbf{x}_l,\textbf{x}_s)d\textbf{x}_{l}.
    \label{PincTE}
  \end{multline}
Finally, we complete our derivation by assigning
  \begin{multline}
  P_{pred}^{scat,TE}(\textbf{x}_u,\textbf{x}_s)= \\
  \int_{\mathcal{V}}\gamma_0^2(\textbf{x})G_{0,tar}(\textbf{x}_u,\textbf{x})\chi^c(\textbf{x})P^{inc,TE}(\textbf{x},\textbf{x}_s)d\textbf{x}.
  \label{PpredTE}
  \end{multline}

Our approach computes an incident wavefield, which contains all of the information from the surrounding medium of the target of interest. This means our approach is not limited to one kind of parametrization, and it can be implemented for any parametrization choice. Further, It is also possible to inject $S_u$ and $S_l$ as dipole sources using a finite-difference algorithm instead of solving Equations \ref{PincTE} and \ref{PpredTE}. Finally, we can solve the following least-squares problem, i.e., minimizing the objective function:
  \begin{equation}
    J(\delta \textbf{m})=\frac{1}{2}\|\textbf{P}_{pred}^{scat,TE}(\delta \textbf{m})-\textbf{P}_{obs}^{scat,TE}\|^2_2.
    \label{costfun_modified}
  \end{equation}

\subsection{Marchenko Green's function retrieval}

  We obtained a target-enclosed LSRTM algorithm in the previous section. Nevertheless, in most real-world situations, one does not have physical access to the boundaries of the region of interest. The alternative to physical receivers inside the medium is to create virtual receivers with redatuming. Marchenko redatuming is a state-of-the-art data-driven approach that can compute Green's functions at any depth level with all orders of multiple reflections from the reflection response at the surface and a smooth background model of the medium.
  
  To summarise, these redatumed Green's functions are retrieved by iteratively solving the Marchenko-type representations. These representations are \cite{Wapenaar6}

  \begin{multline}
    G^-_{Mar}(\textbf{x}_{v},\textbf{x}_{s}) = \int_{{\mathfrak{d}\mathcal{V}_0}} R(\textbf{x}_s,\textbf{x}'_s)f_1^+(\textbf{x}'_s,\textbf{x}_{v}) \,d\textbf{x}'_s  \\
     -f_1^-(\textbf{x}_{s},\textbf{x}_{v}), 
    \label{Mar_r-}
\end{multline}

and

\begin{multline}
    G^+_{Mar}(\textbf{x}_{v},\textbf{x}_{s}) = -\int_{{\mathfrak{d}\mathcal{V}_0}} R(\textbf{x}_s,\textbf{x}'_s)f_1^-(\textbf{x}'_s,\textbf{x}_{v})^* \,d\textbf{x}'_s \\
     + f_1^+(\textbf{x}_{s},\textbf{x}_{v})^* .
    \label{Mar_r+}
\end{multline}

Here, $\mathfrak{d}\mathcal{V}_0$ is the surface, $\textbf{x}_s$ and $\textbf{x}'_s$ are locations at the surface, and $\textbf{x}_v$ is a virtual location on an arbitrary depth. Moreover, $f^\pm_1$ are upgoing (-) and downgoing (+) parts of the focusing function. In addition, $R(\textbf{x}_s,\textbf{x}'_s)$ is the reflection response at the surface which is related to the upgoing Green's function of the medium via

\begin{equation}
    R(\textbf{x}_s,\textbf{x}'_s) = \frac{2\partial'_{3,s}}{-\rho(\textbf{x}'_s)} G^-(\textbf{x}_s,\textbf{x}'_s).
\label{R_to_r}
\end{equation}

Here, $\partial'_{3,s}$ is the vertical partial derivative at $\textbf{x}'_s$. We refer to \cite{Thorbecke2} for a comprehensive explanation of the derivation and numerical algorithms for solving these equations.

Consequently, we can substitute Marchenko Green's functions with the target boundaries Green's functions as follows:

\begin{equation}
    G^+(\textbf{x}_{u},\textbf{x}_{s}) \approx G^+_{Mar}(\textbf{x}_{u},\textbf{x}_{s}),
\label{Mar_Gp}
\end{equation}

\begin{equation}
    G^-(\textbf{x}_{u},\textbf{x}_{s}) \approx G^-_{Mar}(\textbf{x}_{u},\textbf{x}_{s}),
\label{Mar_Gm_up}
\end{equation}

and

\begin{equation}
    G^-(\textbf{x}_{l},\textbf{x}_{s}) \approx G^-_{Mar}(\textbf{x}_{l},\textbf{x}_{s}).
\label{Mar_Gm_low}
\end{equation}

Equations \ref{Mar_Gp} and \ref{Mar_Gm_low} can be used in equations \ref{Source_udf} and \ref{Source_ldf} to obtain the source terms $S_u$ and $S_l$, whereas equation \ref{Mar_Gm_up} can be used in equation \ref{PobsTE} to obtain the observed target-oriented scattered response.

\section{\label{sec:3} Numerical results}

\subsection{\label{sec:3.1} single-sided algorithm vs. double-sided algorithm}

\begin{figure}
\begin{subfigure}{0.5\textwidth}
\includegraphics[width=\textwidth]{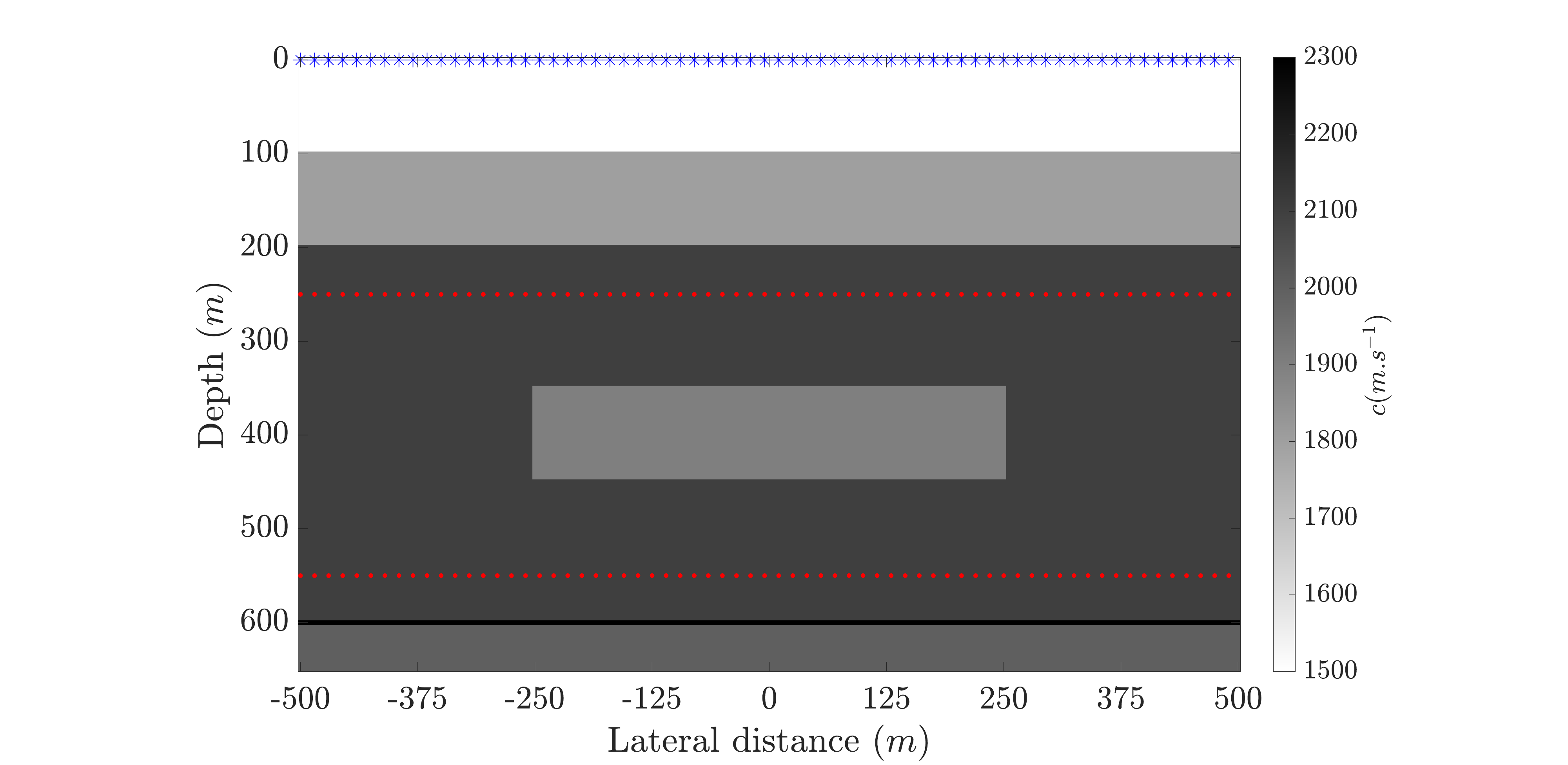}
\caption{}
\label{velocity_model}
\end{subfigure}
\hfill
\begin{subfigure}{0.5\textwidth}
\includegraphics[width=\textwidth]{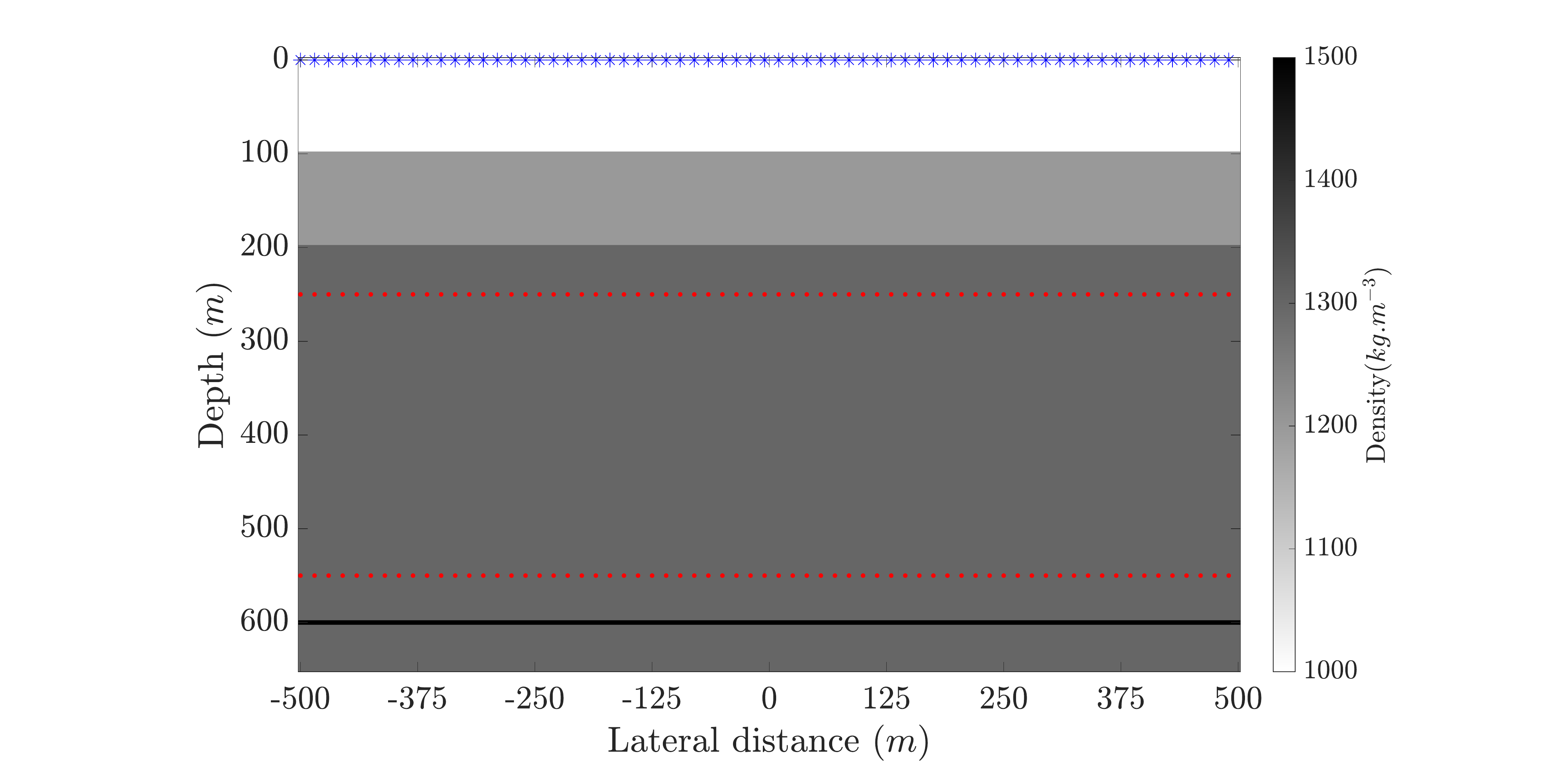}
\caption{}
\label{density_model}
\end{subfigure}
\caption{(a) Velocity model and (b) density model. The blue stars at the surface are source locations, and the red dots are the boundaries of the target.}
\label{model}
\end{figure}

\begin{figure}
\centering
\includegraphics[width=0.4\textwidth]{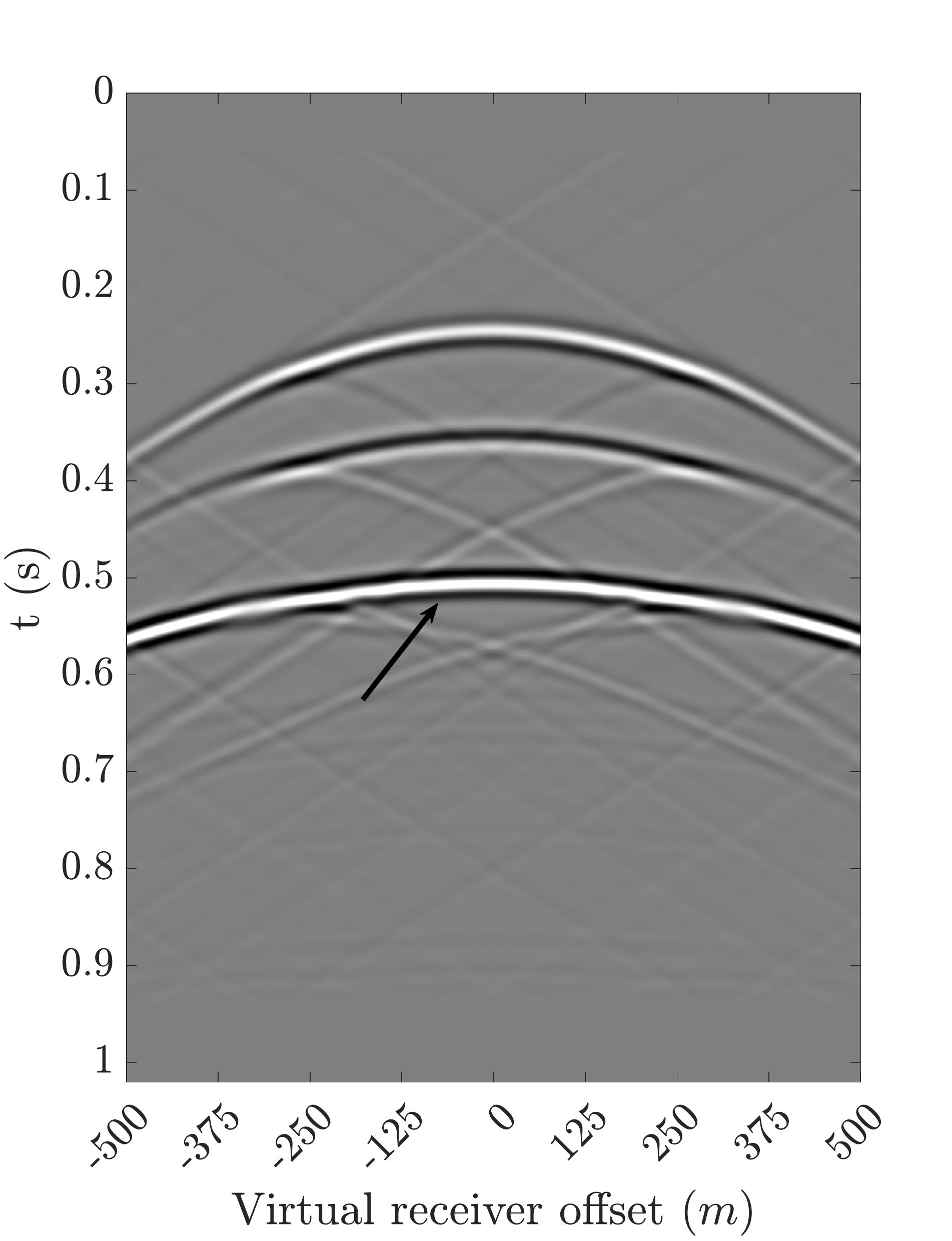}
\caption{Upgoing component of data at the upper boundary with a source located at $\textbf{x}_s=(0,0)$. This is the data corresponding to the first term on the right-hand side of Equation \ref{PobsTE}.}
\label{obs}
\end{figure}

In this section, we aim to visualize the performance of target-enclosed LSRTM. Here we use direct modeling of the Green's functions with receivers inside the medium; in section B we use the Marchenko method to retrieve these Green's functions from the reflection response at the surface. A model with dimensions of 1000 $m$ by 650 $m$ is designed as shown in Fig. \ref{model}. The spatial grid sampling is 5 $m$ in both directions. The target region consists of a rectangular velocity anomaly embedded in a homogeneous background and a constant density. A total of 201 sources are placed at the surface, and 402 receivers are on the target's upper (250 $m$) and lower (550 $m$) boundaries. The required Green's functions and wavefields are computed by a finite-difference algorithm \cite{Thorbecke1} and a Ricker wavelet with a dominant frequency of 30 $Hz$, where the recording time sampling of the receivers is set to 4 $ms$. According to the theory section, the wavefields at the receiver positions are decomposed into upgoing and downgoing components. Fig. \ref{obs} shows the upgoing component of the data at the upper boundary with a source located at $\textbf{x}_s=(0,0)$, corresponding to the first term on the right-hand side of Equation \ref{PobsTE}. We study the effects of including the lower boundary with two different cases: 1) a homogeneous background velocity and 2) a smooth background velocity for migration.

\subsubsection{\label{sec:3.1.1} Homogeneous background velocity}

\begin{figure}
\centering
\includegraphics[width=0.5\textwidth]{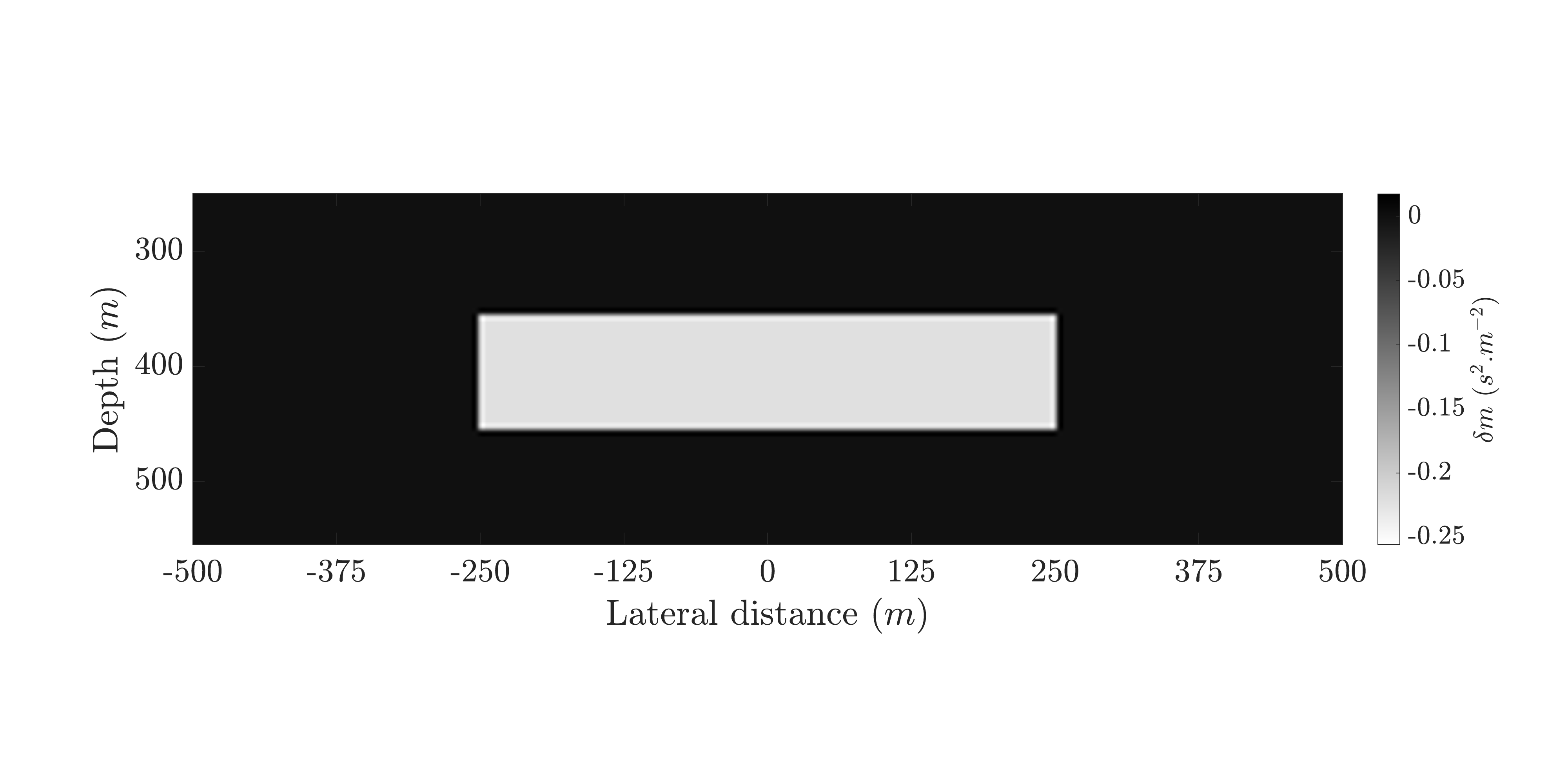}
\caption{True perturbation model in a homogeneous background.}
\label{pert_homogen}
\end{figure}

\begin{figure}
\centering
\begin{subfigure}{0.5\textwidth}
\centering
\includegraphics[width=0.5\textwidth]{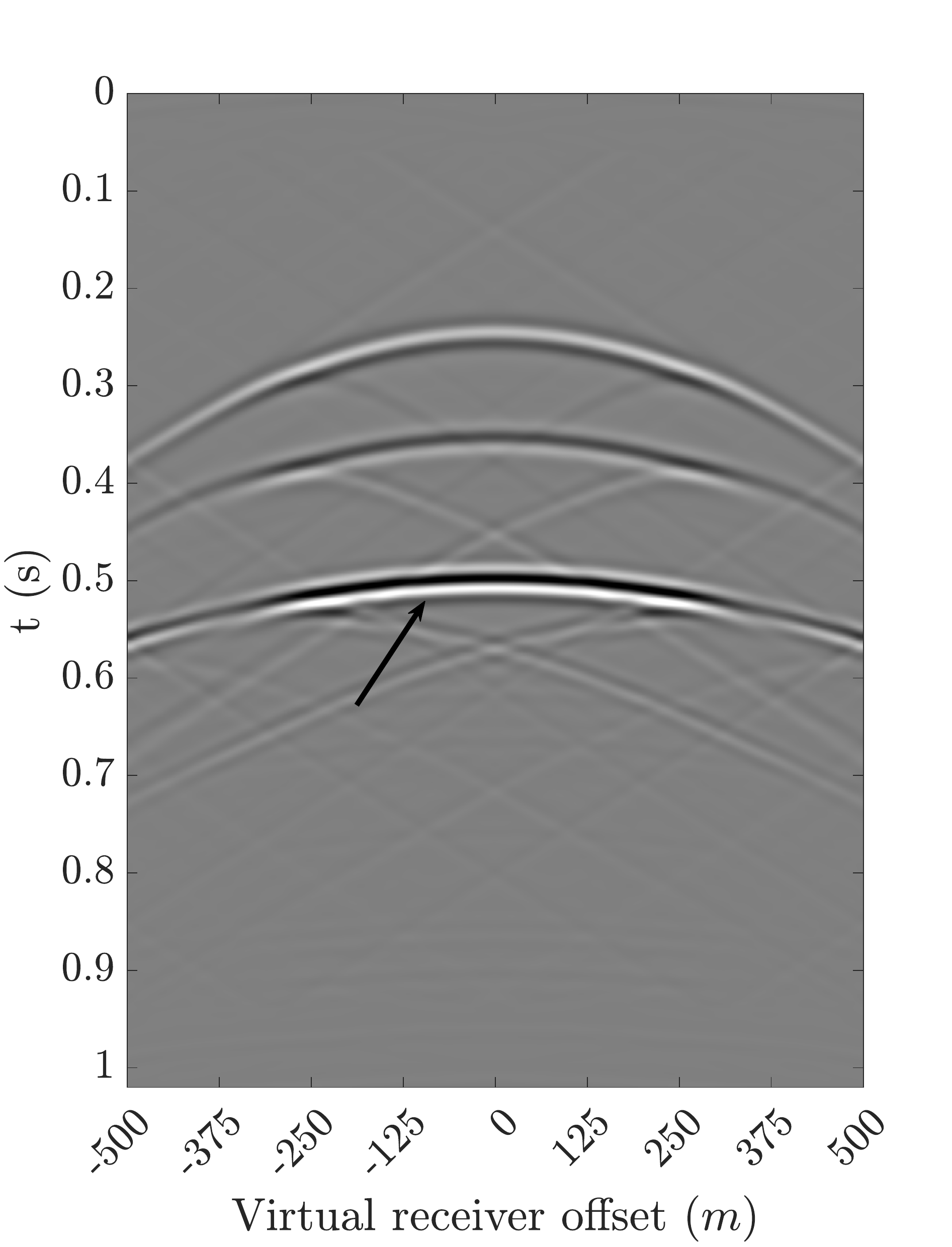}
\caption{}
\label{obs_enc_homogen}
\end{subfigure}
\hfill
\begin{subfigure}{0.5\textwidth}
\centering
\includegraphics[width=0.5\textwidth]{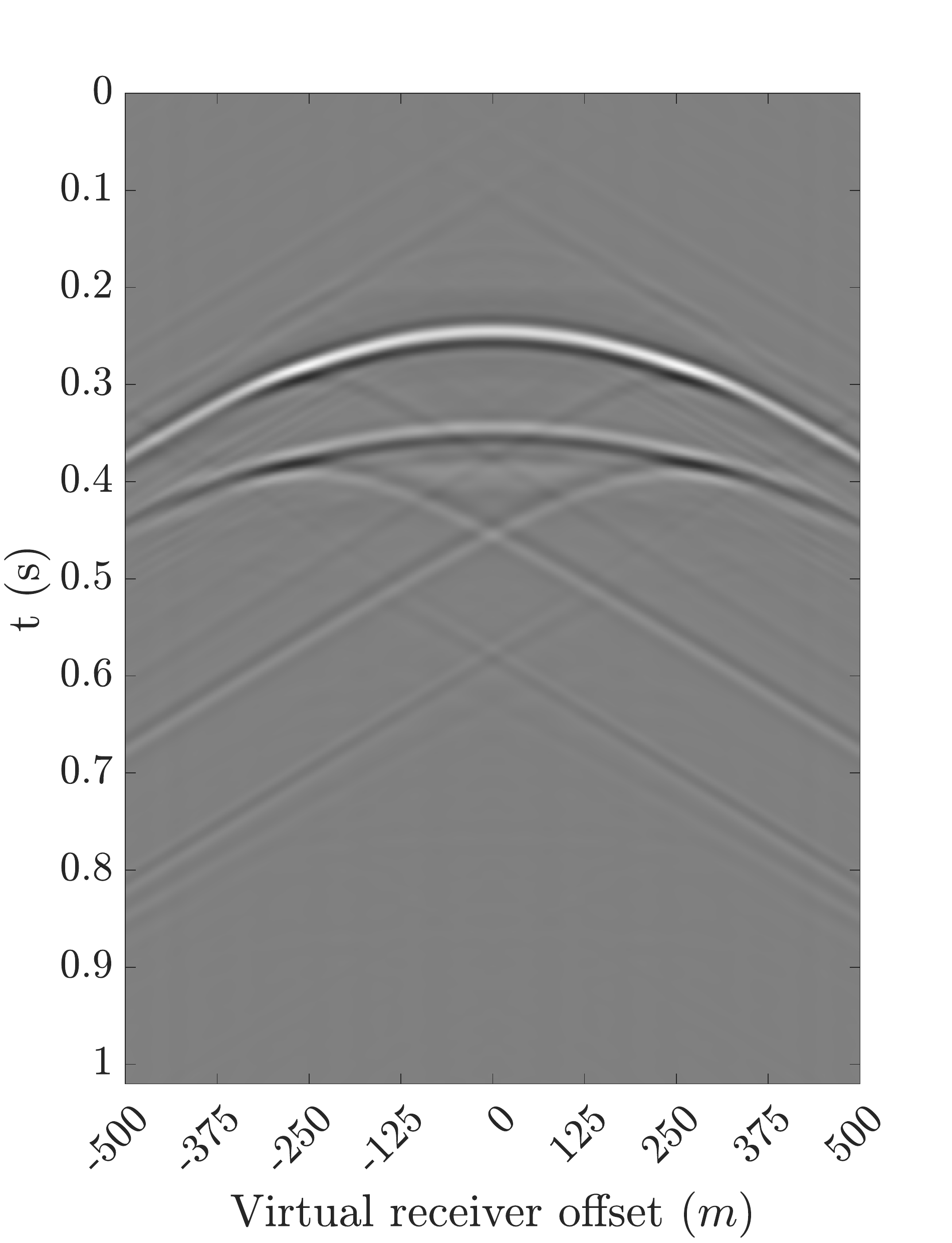}
\caption{}
\label{pred_or_homogen}
\end{subfigure}
\hfill
\begin{subfigure}{0.5\textwidth}
\centering
\includegraphics[width=0.5\textwidth]{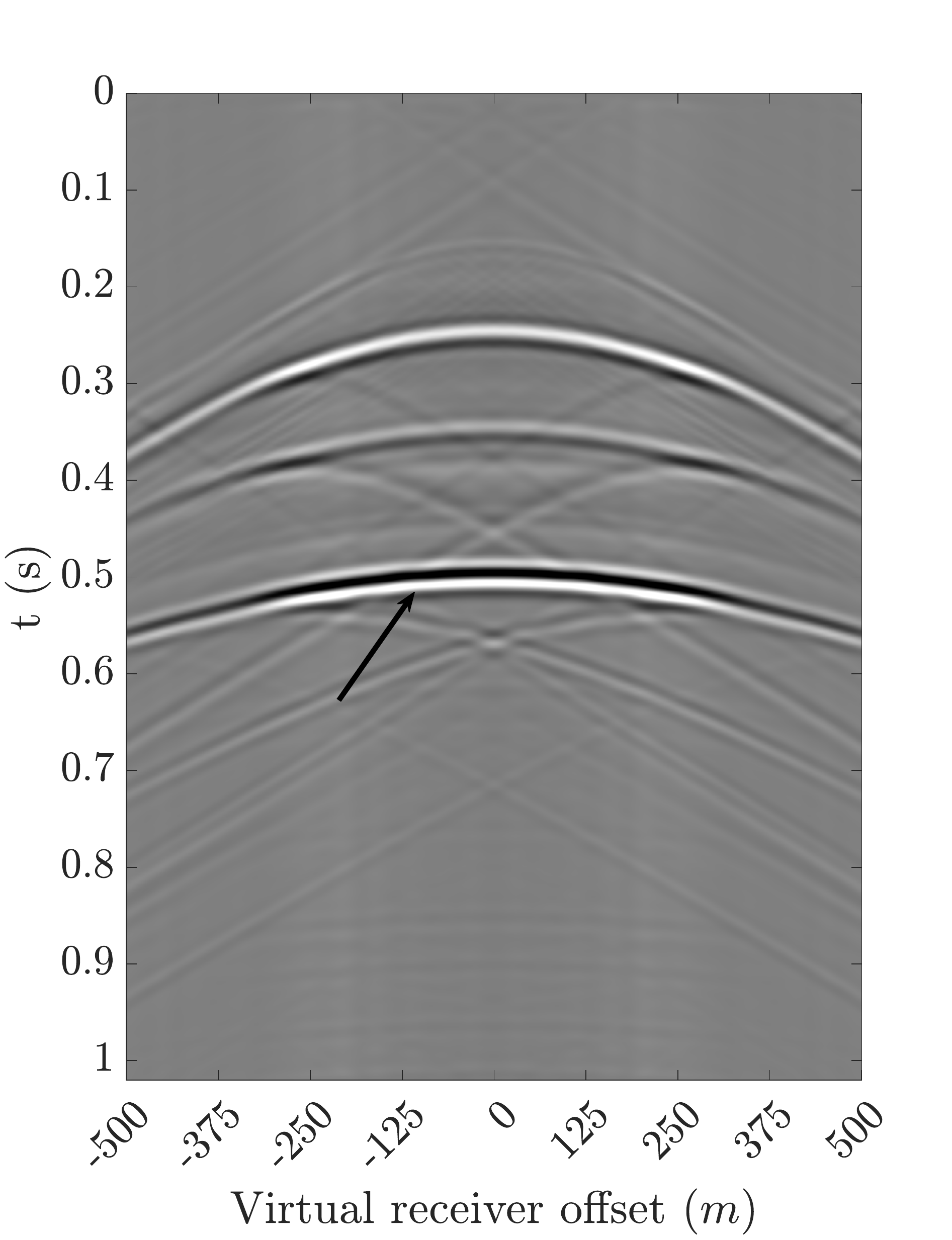}
\caption{}
\label{pred_enc_homogen}
\end{subfigure}
\caption{Homogeneous background velocity case, data domain: (a) Upgoing component of double-sided observed data at the upper boundary (Equation \ref{PobsTE}), (b) predicted data of single-sided algorithm after 30 iterations of LSRTM, and (c) predicted data of double-sided algorithm after 30 iterations of LSRTM. All wavefields are recorded at the upper boundary of the target (250 $m$).}
\label{data_homogen}
\end{figure}

In this section, we show the results using a homogeneous background velocity in the target area. The true perturbation model of this case is shown in Fig. \ref{pert_homogen}. We design two scenarios to demonstrate the performance and consequences of the target-enclosed algorithm. For the first scenario, we only consider the upper boundary of the target in the inversion process and completely ignore the lower boundary contribution. We call this scenario "single-sided algorithm". For this scenario, the observed data is the same as Fig. \ref{obs}. For the second one, we include the lower boundary contribution, which is the "double-sided algorithm" explained in the previous section. The observed data for this scenario is shown in Fig. \ref{obs_enc_homogen}, which corresponds to the left-hand side of Equation \ref{PobsTE}, whereas Fig. \ref{obs} shows $G^-(\textbf{x}_u,\textbf{x}_s)W(\omega)$. The black arrows in Figures \ref{obs} and \ref{obs_enc_homogen} indicate the full reflection from the reflector below the lower boundary and the forward scattered part of it, respectively. Figures \ref{pred_or_homogen} and \ref{pred_enc_homogen} show the predicted data of single-sided and double-sided algorithms, respectively, after 30 iterations of LSRTM. A detailed investigation of Fig. \ref{data_homogen} proves that the double-sided algorithm can predict the forward scattered event that is passing through the perturbation which is indicated by a black arrow in Fig. \ref{pred_enc_homogen}.  

\begin{figure}
\centering
\begin{subfigure}{0.5\textwidth}
\centering
\includegraphics[width=0.8\textwidth]{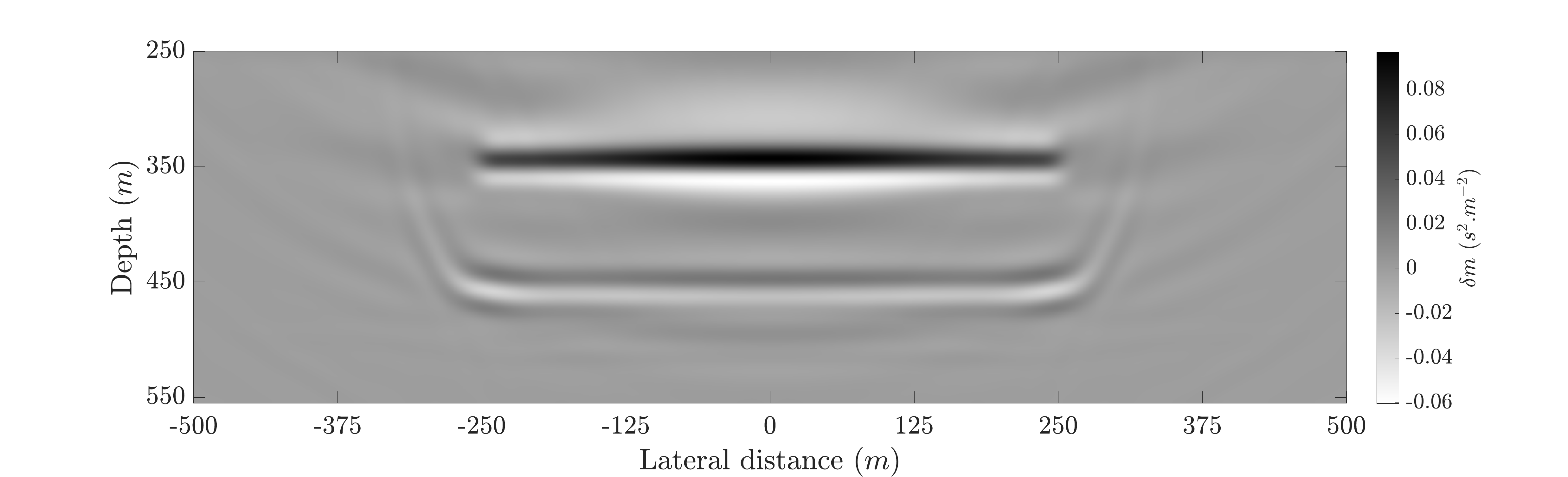}
\caption{}
\label{RTM_or_homogen}
\end{subfigure}
\hfill
\begin{subfigure}{0.5\textwidth}
\centering
\includegraphics[width=0.8\textwidth]{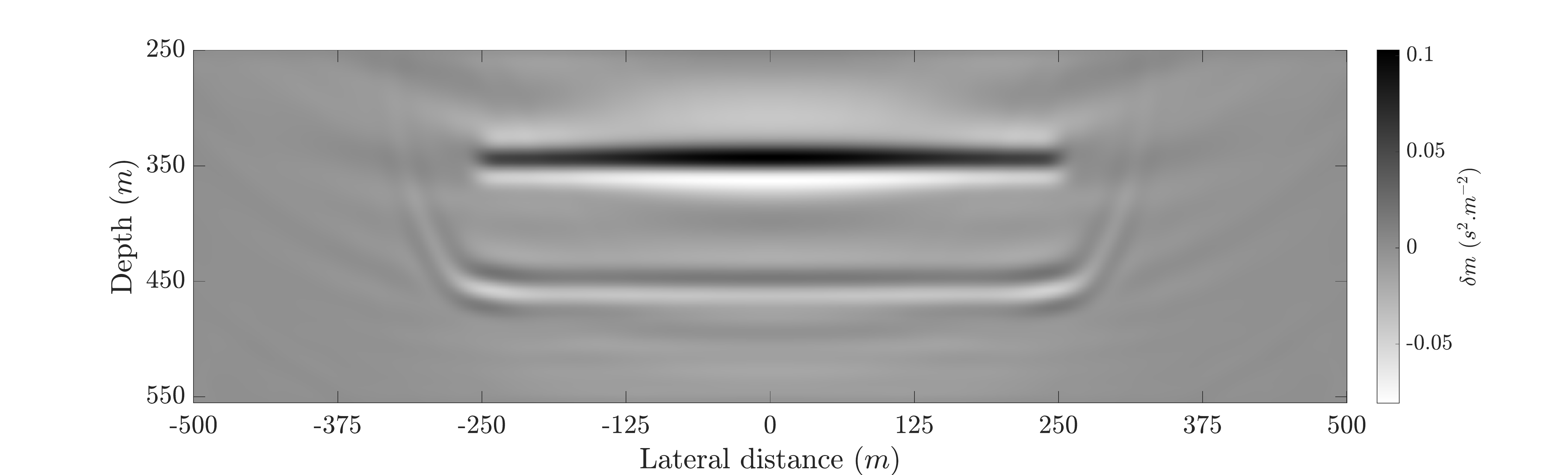}
\caption{}
\label{RTM_Enc_homogen}
\end{subfigure}
\hfill
\begin{subfigure}{0.5\textwidth}
\centering
\includegraphics[width=0.8\textwidth]{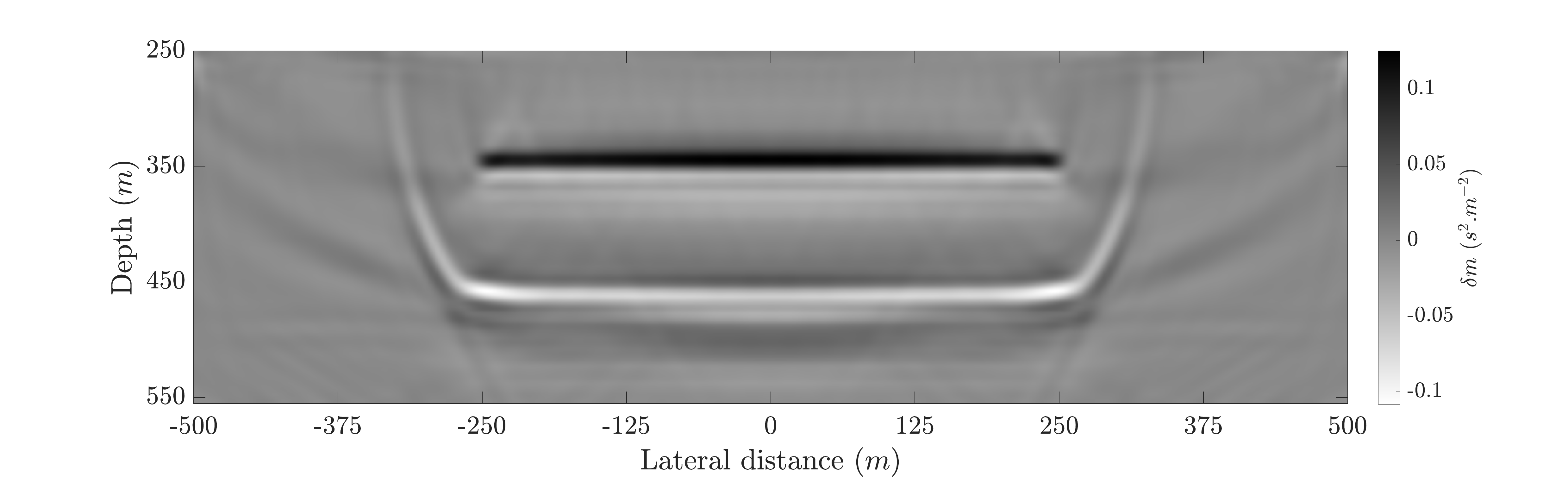}
\caption{}
\label{LSRTM_or_homogen}
\end{subfigure}
\hfill
\begin{subfigure}{0.5\textwidth}
\centering
\includegraphics[width=0.8\textwidth]{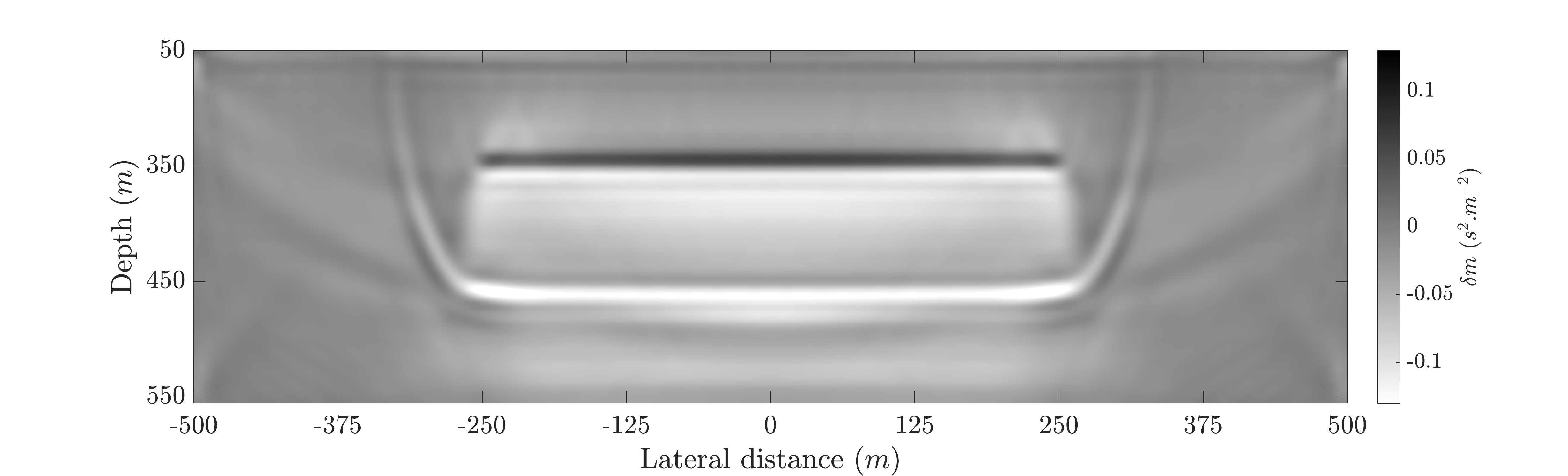}
\caption{}
\label{LSRTM_Enc_homogen}
\end{subfigure}
\caption{Homogeneous background velocity case, image domain: (a) RTM image of the single-sided algorithm, (b) RTM image of the double-sided algorithm, (c) LSRTM image of the single-sided algorithm after 30 iterations, and (d) LSRTM image of the double-sided algorithm after 30 iterations.}
\label{img_homogen}
\end{figure}

\begin{figure}
\centering
\includegraphics[width=0.5\textwidth]{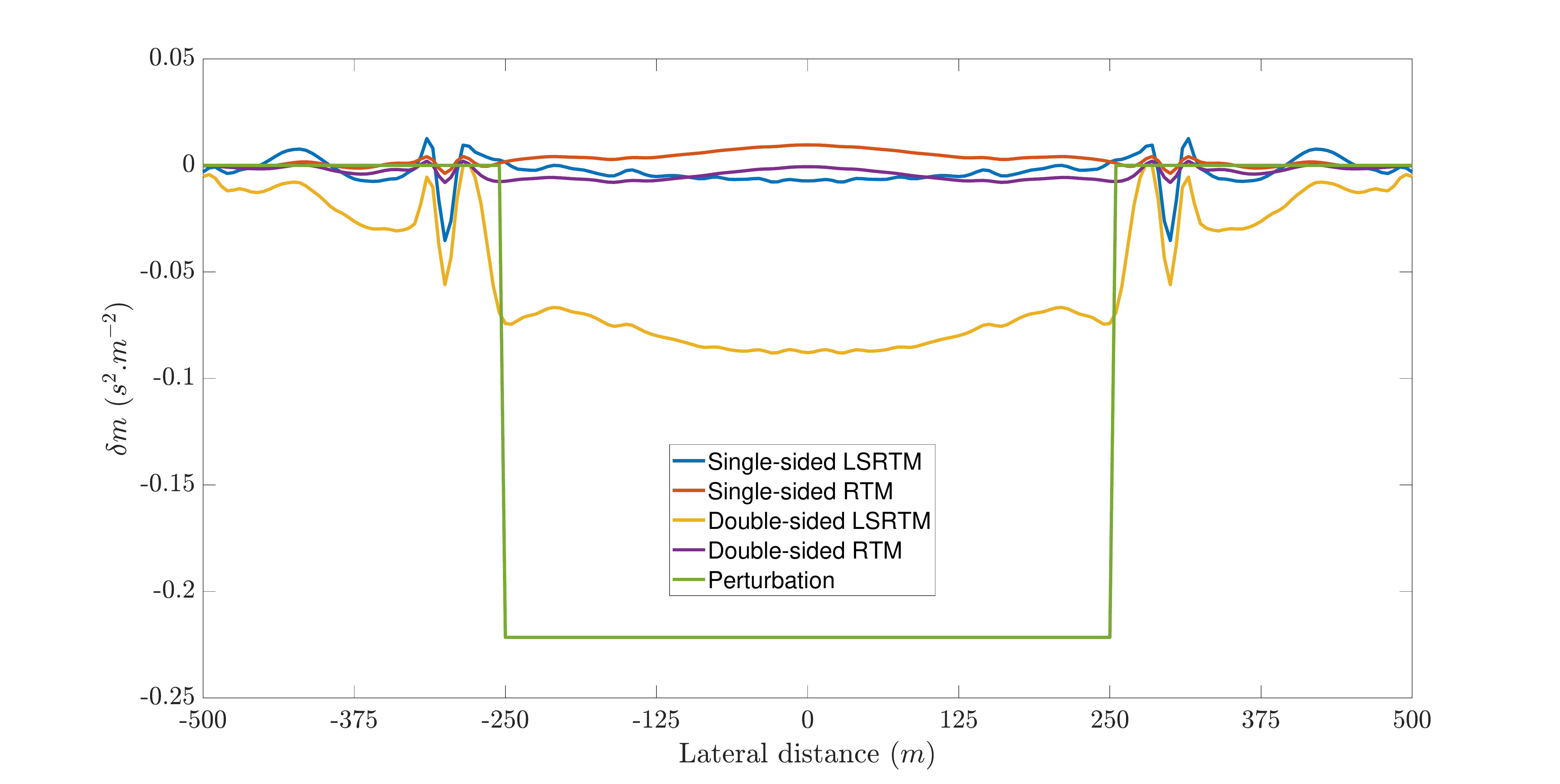}
\caption{Horizontal cross-section at the depth of 400 m of the retrieved perturbation with a homogeneous migration velocity.}
\label{horiz_homogen}
\end{figure}

To move our investigation further, we show the imaging results of both algorithms in Fig. \ref{img_homogen}. Figures \ref{RTM_or_homogen} and \ref{RTM_Enc_homogen} show the RTM images of both approaches. As we can see, the RTM result of the double-sided algorithm faintly reveals the long wavelength part of the model. Moving to the LSRTM results in Figures \ref{LSRTM_or_homogen} and \ref{LSRTM_Enc_homogen}, we observe an interesting outcome. The LSRTM result of the single-sided algorithm shows that it cannot recover the long wavelength part of the model (Fig. \ref{LSRTM_or_homogen}). In contrast, the double-sided algorithm can incorporate the information embedded inside the forward scattered field (Fig. \ref{LSRTM_Enc_homogen}), and it recovers the long wavelength parts of the volume perturbation. Moreover, Fig. \ref{horiz_homogen} shows the horizontal cross-section of the retrieved perturbation. In this figure, we can see the double-sided approach can recover the vertical boundaries of the perturbation. Nevertheless, since the background velocity for the migration is not updated during LSRTM, the fit of the reflected event from below the target is not accurate.

\begin{figure}
\centering
\includegraphics[width=0.5\textwidth]{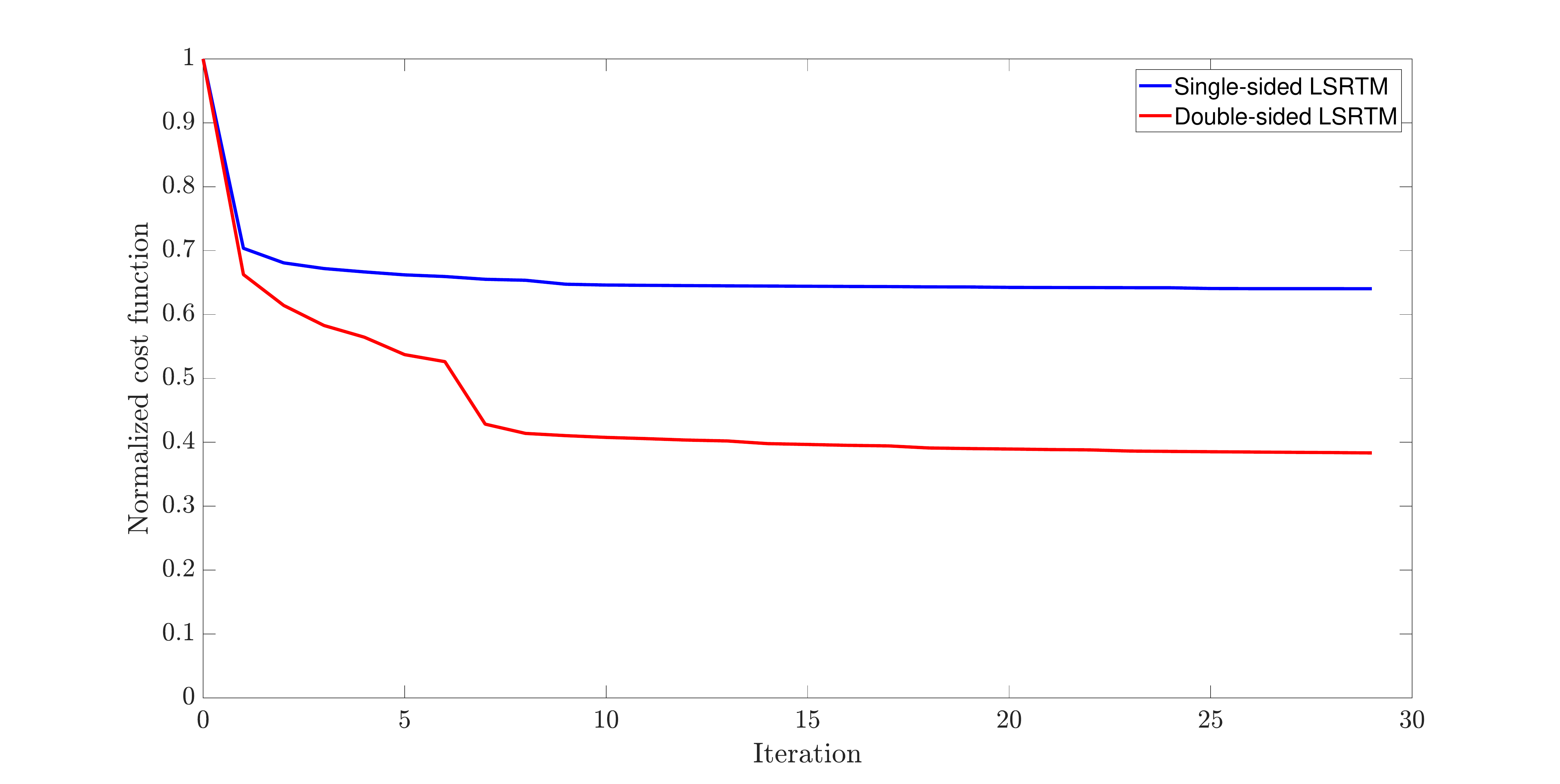}
\caption{Cost function comparison of homogeneous background velocity case.}
\label{cost_homogen}
\end{figure}

To conclude this section, we compare the cost functions of both approaches in Fig. \ref{cost_homogen}. The cost function of target-oriented LSRTM shows a slow convergence rate. In comparison, the target-enclosed approach includes the extra information coming from the lower boundary, so its cost function converges faster and to a lower minimum. 

\subsubsection{\label{sec:3.1.2} Smooth background velocity}

\begin{figure}
\centering
\includegraphics[width=0.5\textwidth]{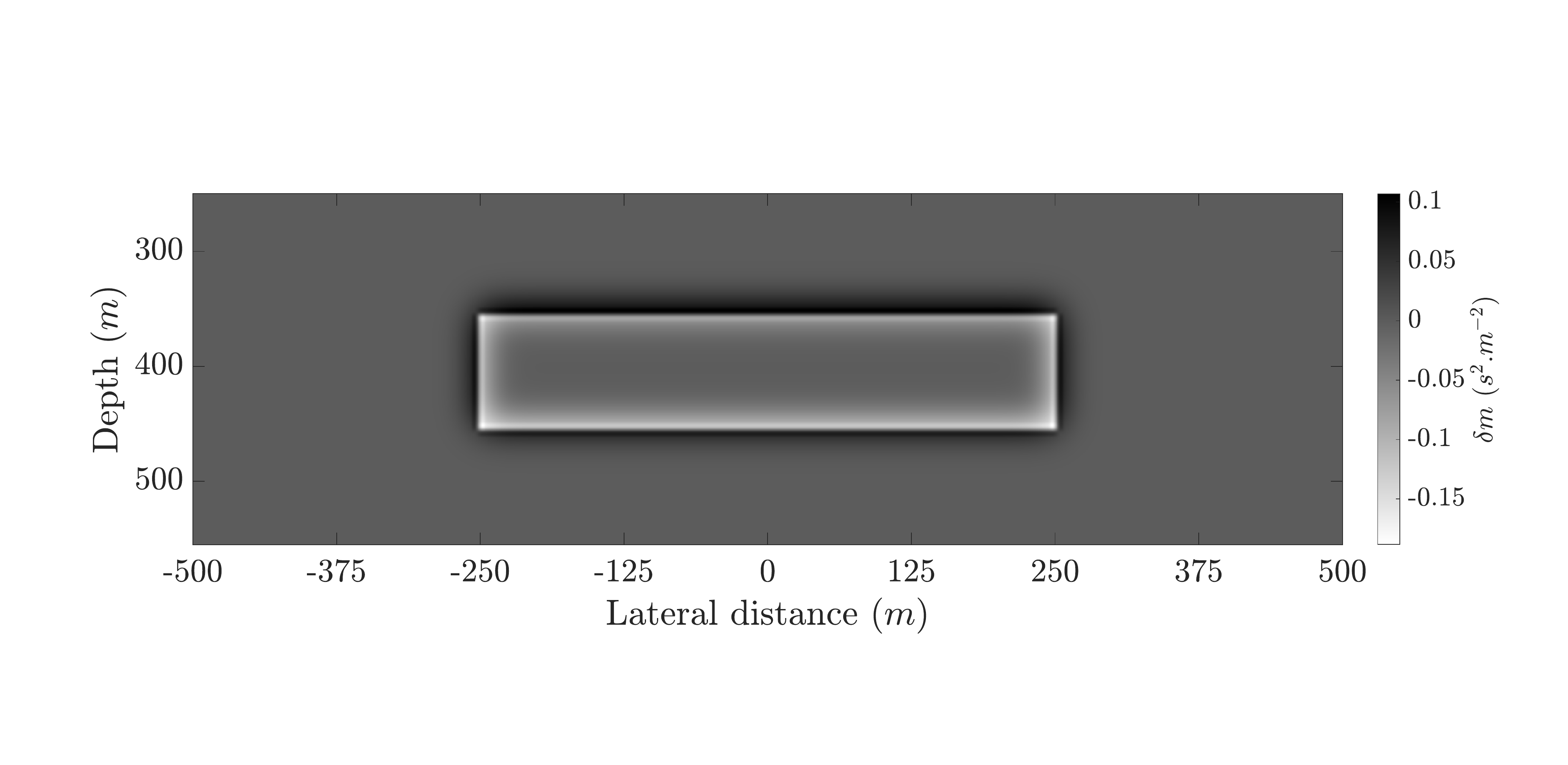}
\caption{True perturbation model in a smooth background.}
\label{pert_smooth}
\end{figure}

\begin{figure}
\centering
\begin{subfigure}{0.5\textwidth}
\centering
\includegraphics[width=0.5\textwidth]{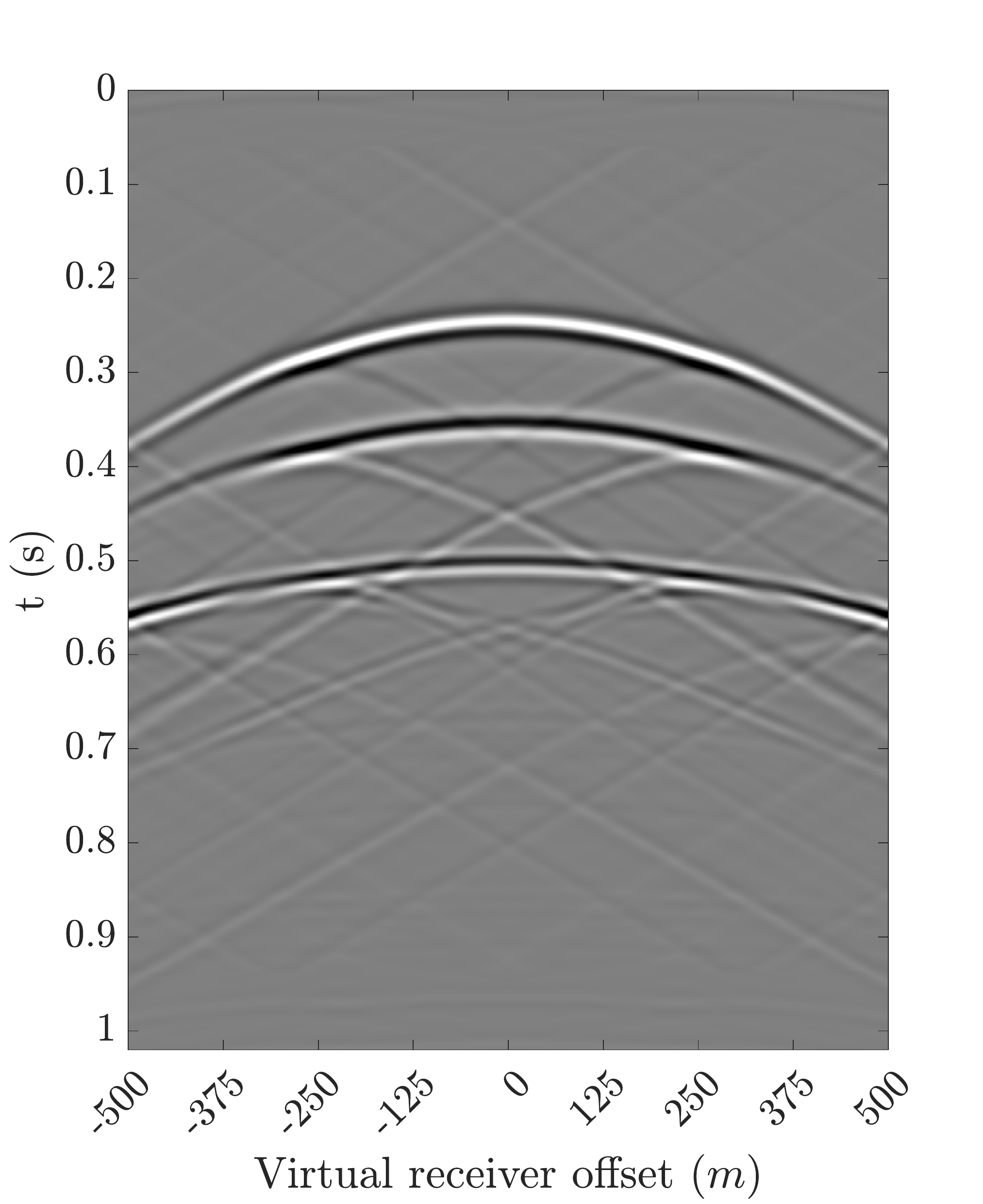}
\caption{}
\label{obs_enc_smooth}
\end{subfigure}
\hfill
\begin{subfigure}{0.5\textwidth}
\centering
\includegraphics[width=0.5\textwidth]{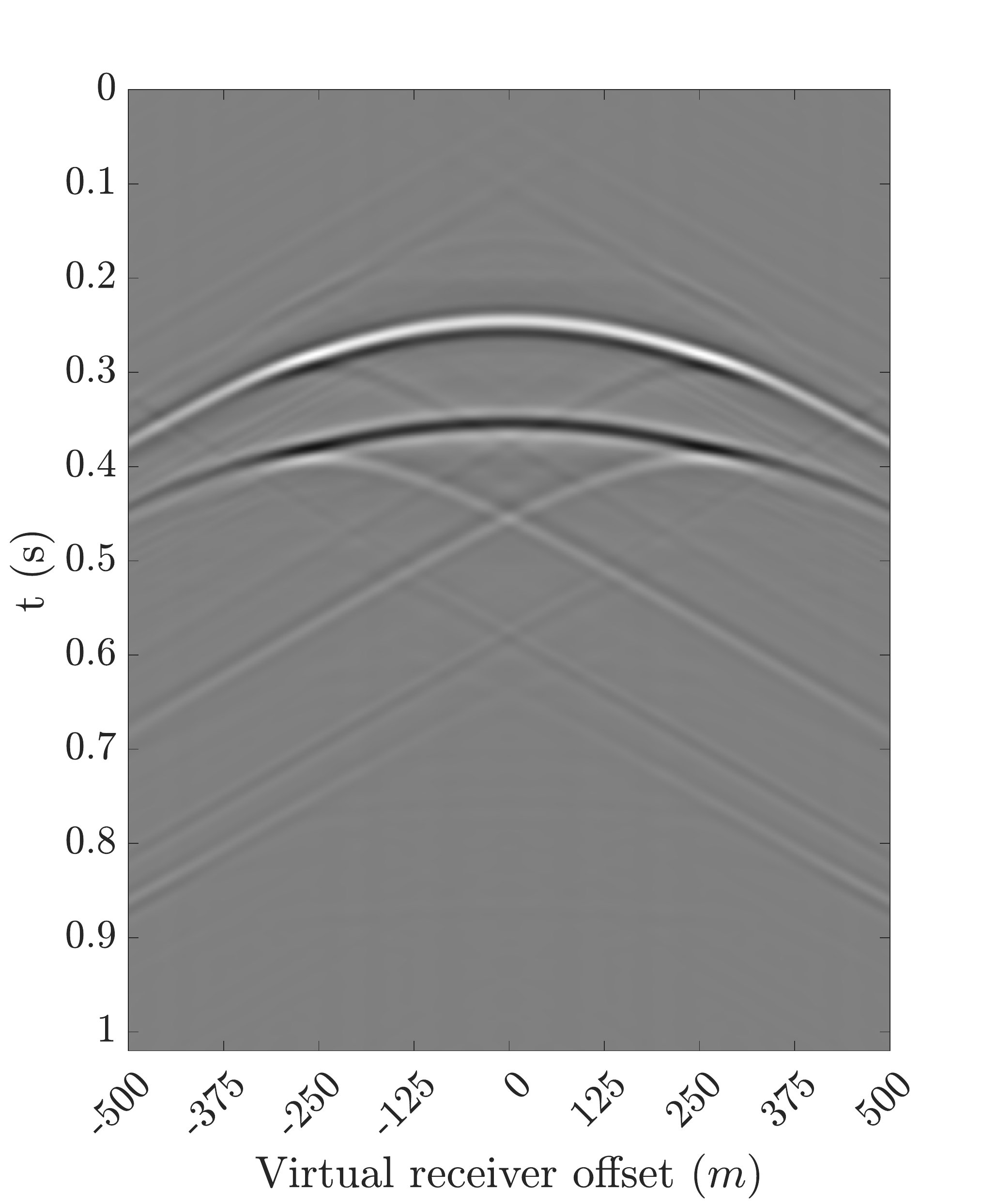}
\caption{}
\label{pred_or_smooth}
\end{subfigure}
\hfill
\begin{subfigure}{0.5\textwidth}
\centering
\includegraphics[width=0.5\textwidth]{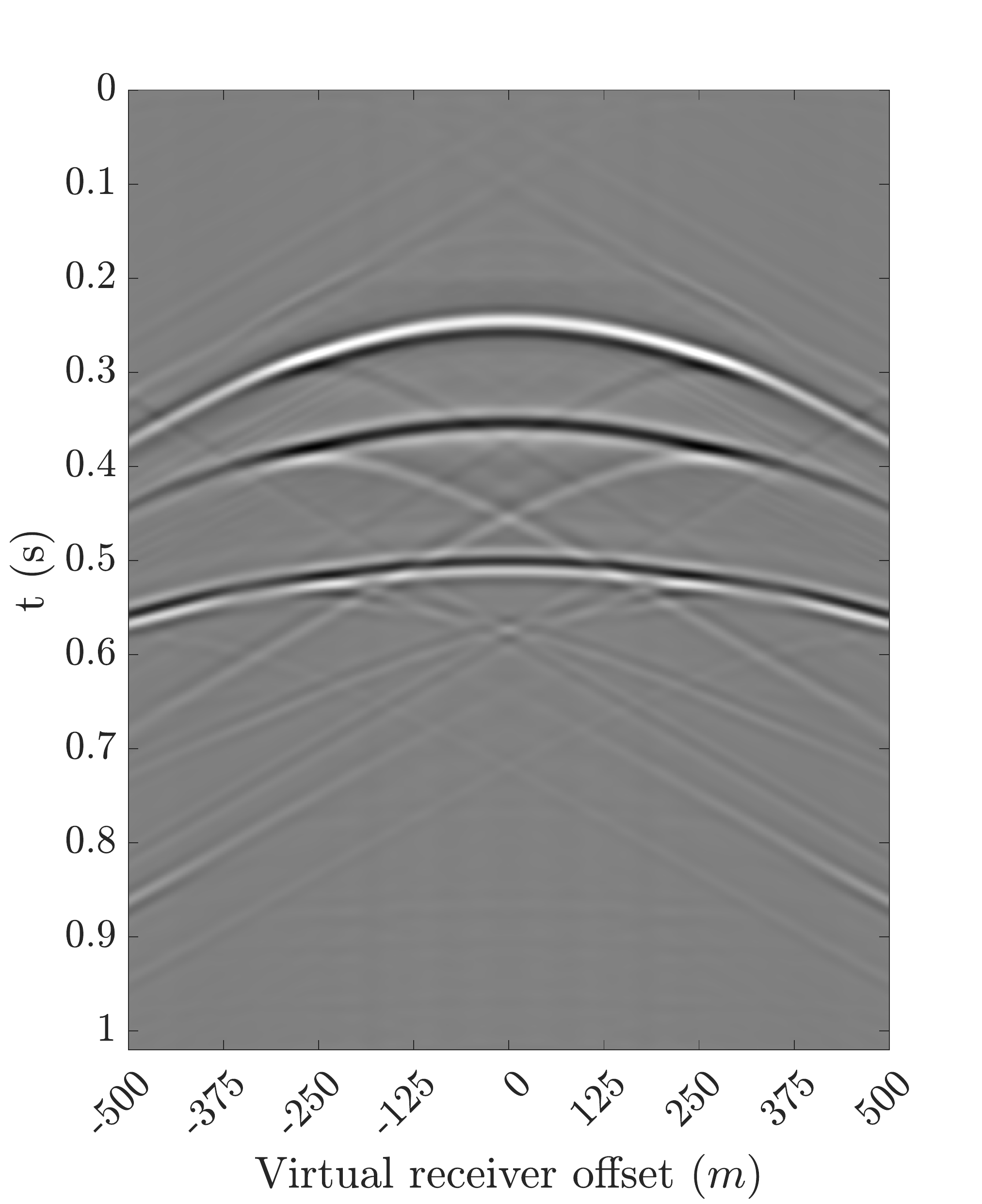}
\caption{}
\label{pred_enc_smooth}
\end{subfigure}
\caption{Smooth background velocity case, data domain: (a) Upgoing component of double-sided observed data at the upper boundary (Equation \ref{PobsTE}), (b) predicted data of single-sided algorithm after 30 iterations of LSRTM, and (c) predicted data of double-sided algorithm after 30 iterations of LSRTM. All wavefields are recorded at the upper boundary of the target (250 $m$).}
\label{data_smooth}
\end{figure}

This section uses a smooth background velocity for migration. The setup is exactly the same as before, except for the background velocity. The perturbation model for this case is shown in Fig. \ref{pert_smooth}. Again, we do the same two scenarios as before, i.e. "single-sided" and "double-sided." For the single-sided scenario, the observed data is the same as before (Fig. \ref{obs}). However, since the right-hand side of the equation \ref{PobsTE} is computed in a different background velocity, the last primary in the observed data for the double-sided approach (Fig. \ref{obs_enc_smooth}) is slightly different. 

\begin{figure}
\centering
\begin{subfigure}{0.5\textwidth}
\centering
\includegraphics[width=0.8\textwidth]{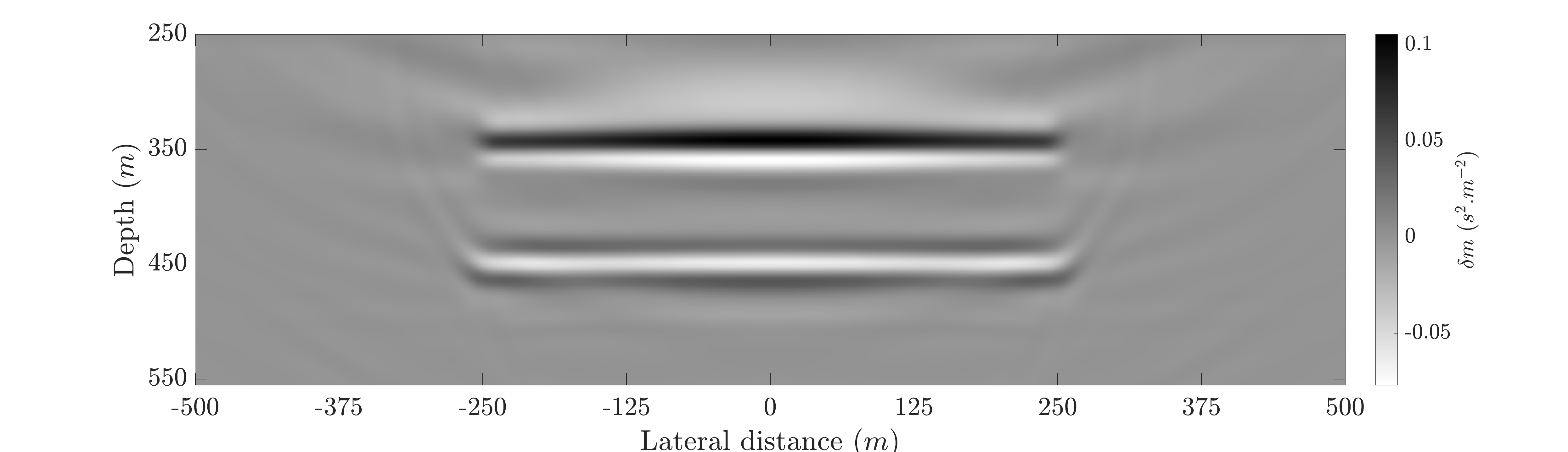}
\caption{}
\label{RTM_or_smooth}
\end{subfigure}
\hfill
\begin{subfigure}{0.5\textwidth}
\centering
\includegraphics[width=0.8\textwidth]{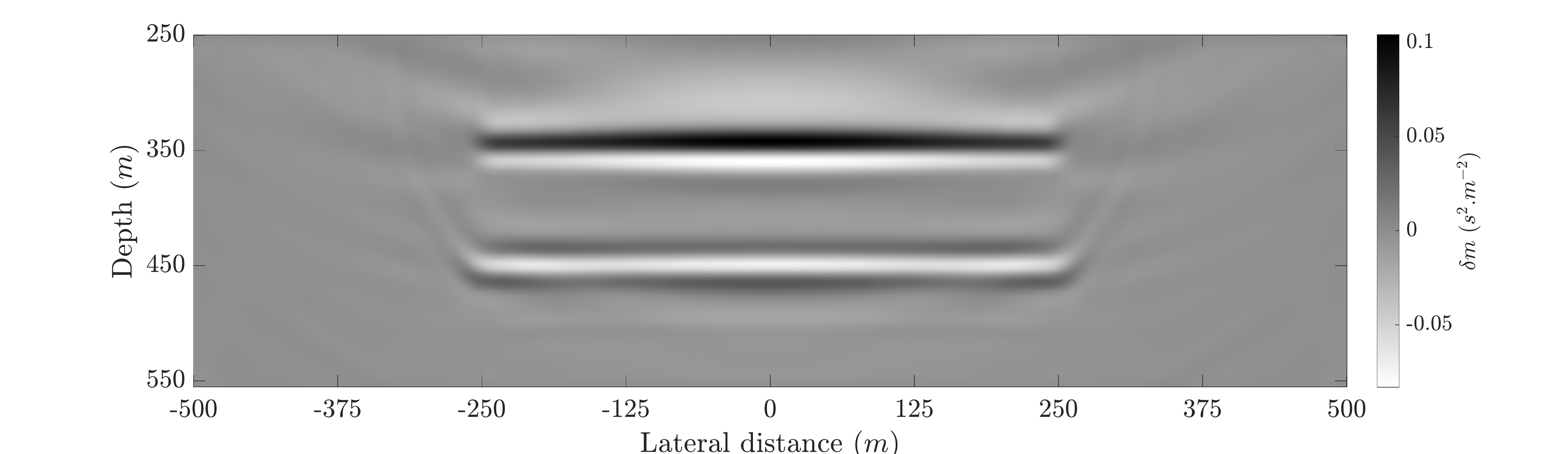}
\caption{}
\label{RTM_Enc_smooth}
\end{subfigure}
\hfill
\begin{subfigure}{0.5\textwidth}
\centering
\includegraphics[width=0.8\textwidth]{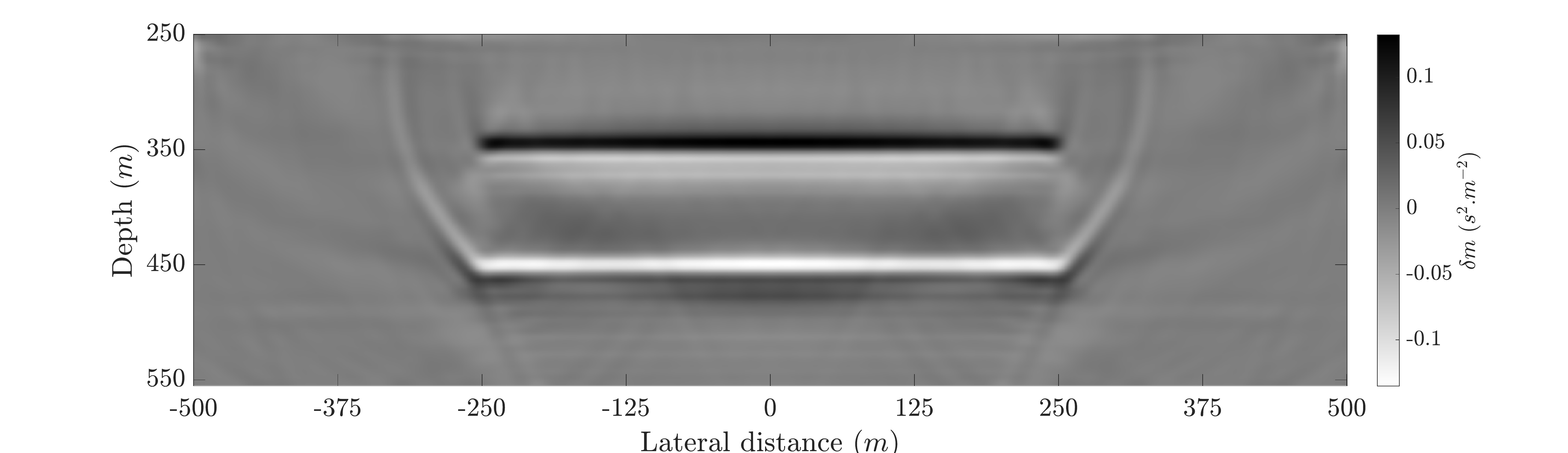}
\caption{}
\label{LSRTM_or_smooth}
\end{subfigure}
\hfill
\begin{subfigure}{0.5\textwidth}
\centering
\includegraphics[width=0.8\textwidth]{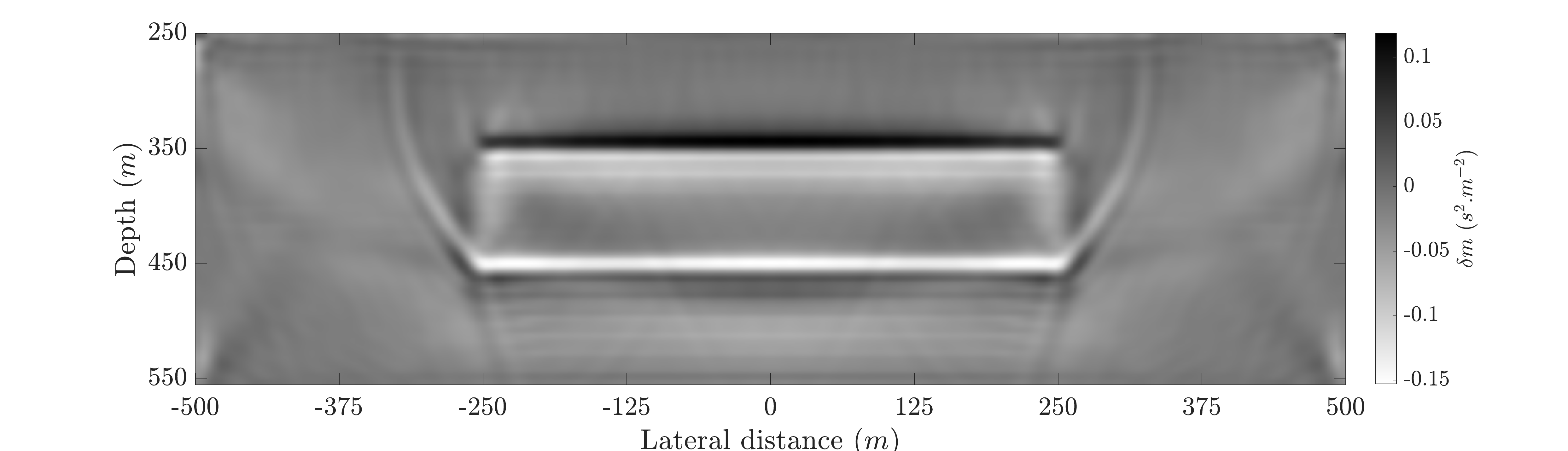}
\caption{}
\label{LSRTM_Enc_smooth}
\end{subfigure}
\caption{Smooth background velocity case, image domain: (a) RTM image of the single-sided algorithm, (b) RTM image of the double-sided algorithm, (c) LSRTM image of the single-sided algorithm after 30 iterations, and (d) LSRTM image of the double algorithm after 30 iterations.}
\label{img_smooth}
\end{figure}

\begin{figure}
\centering
\includegraphics[width=0.5\textwidth]{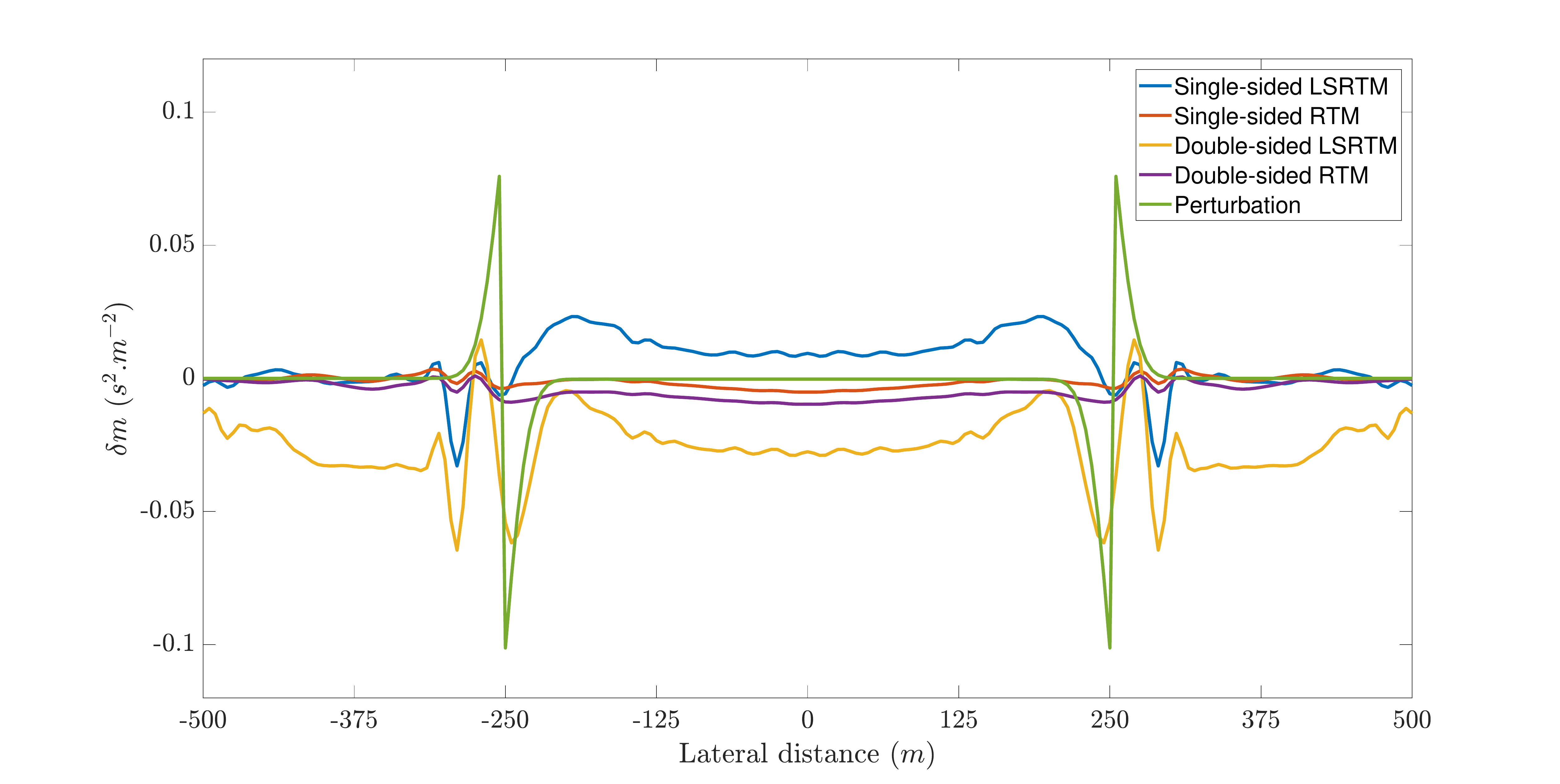}
\caption{Horizontal cross-section at the depth of 400 m of the retrieved perturbation with a smooth migration velocity.}
\label{horiz_smooth}
\end{figure}

\begin{figure}
\centering
\includegraphics[width=0.5\textwidth]{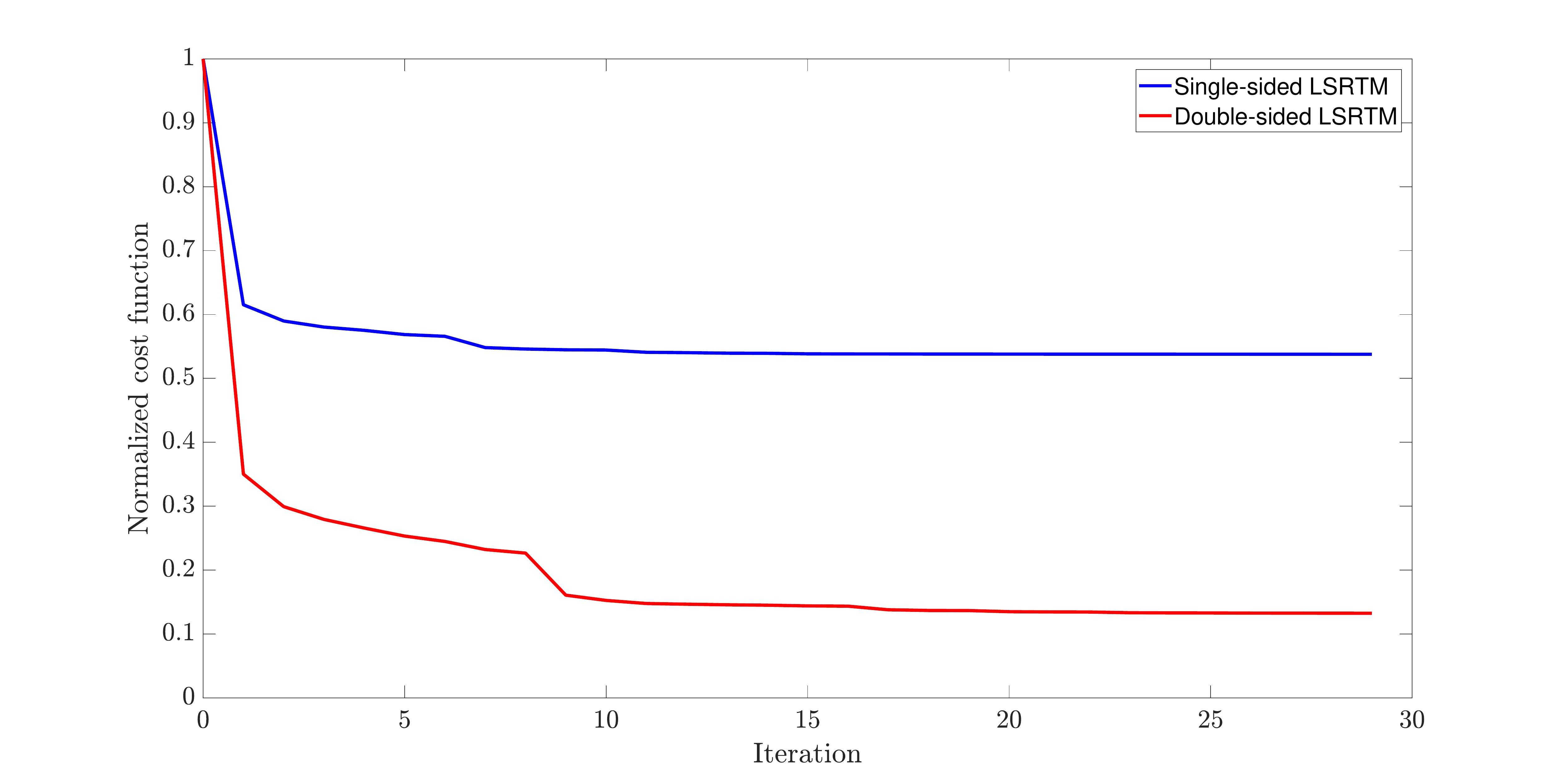}
\caption{Cost function comparison of smooth background velocity case.}
\label{cost_smooth}
\end{figure}

Comparing the results of both approaches shows (Fig. \ref{data_smooth}) that the double-sided approach (Fig. \ref{pred_enc_smooth}) can predict the reflected event coming from the lower boundary. Since we use a smooth background velocity here, this prediction is more accurate than the previous section's results (Fig. \ref{pred_enc_homogen}). In the image domain (Fig. \ref{img_smooth}), the double-sided approach recovers a faint image of the vertical sides of the rectangular perturbation (Fig. \ref{LSRTM_Enc_smooth}) whereas the single-sided results in a more standard image (Fig. \ref{LSRTM_or_smooth}). Moreover, Fig. \ref{horiz_smooth} shows the horizontal cross-section of the retrieved perturbation. In this figure, we can see the double-sided approach can recover the vertical boundaries of the perturbation. The double-sided image is more comparable to the true perturbation in Fig. \ref{pert_smooth}. Finally, investigating the cost functions (Fig. \ref{cost_smooth}) of these approaches shows that the double-sided approach converges better since it can predict the event coming from below the target.

\subsection{\label{sec:3.2} Virtual receivers}

Here, we use the same setup as before but replace the Green's functions at the boundaries with their Marchenko counterparts. In other words, we create virtual receivers with the help of Marchenko redatuming. For a study about the benefits of using Marchenko redatuming instead of a more conventional redatuming algorithm for target-oriented LSRTM we refer to \cite{Shoja3}. In this section, we only focus on including the lower boundary by utilizing virtual receivers created by Marchenko redatuming, and instead of using "double-sided", we use the "target-enclosed" term. Moreover, we show the results for both homogeneous and smooth background velocities in this section.

\subsubsection{\label{sec:3.2.1} Homogeneous background velocity}
\begin{figure}
    \centering
    \includegraphics[width=0.5\textwidth]{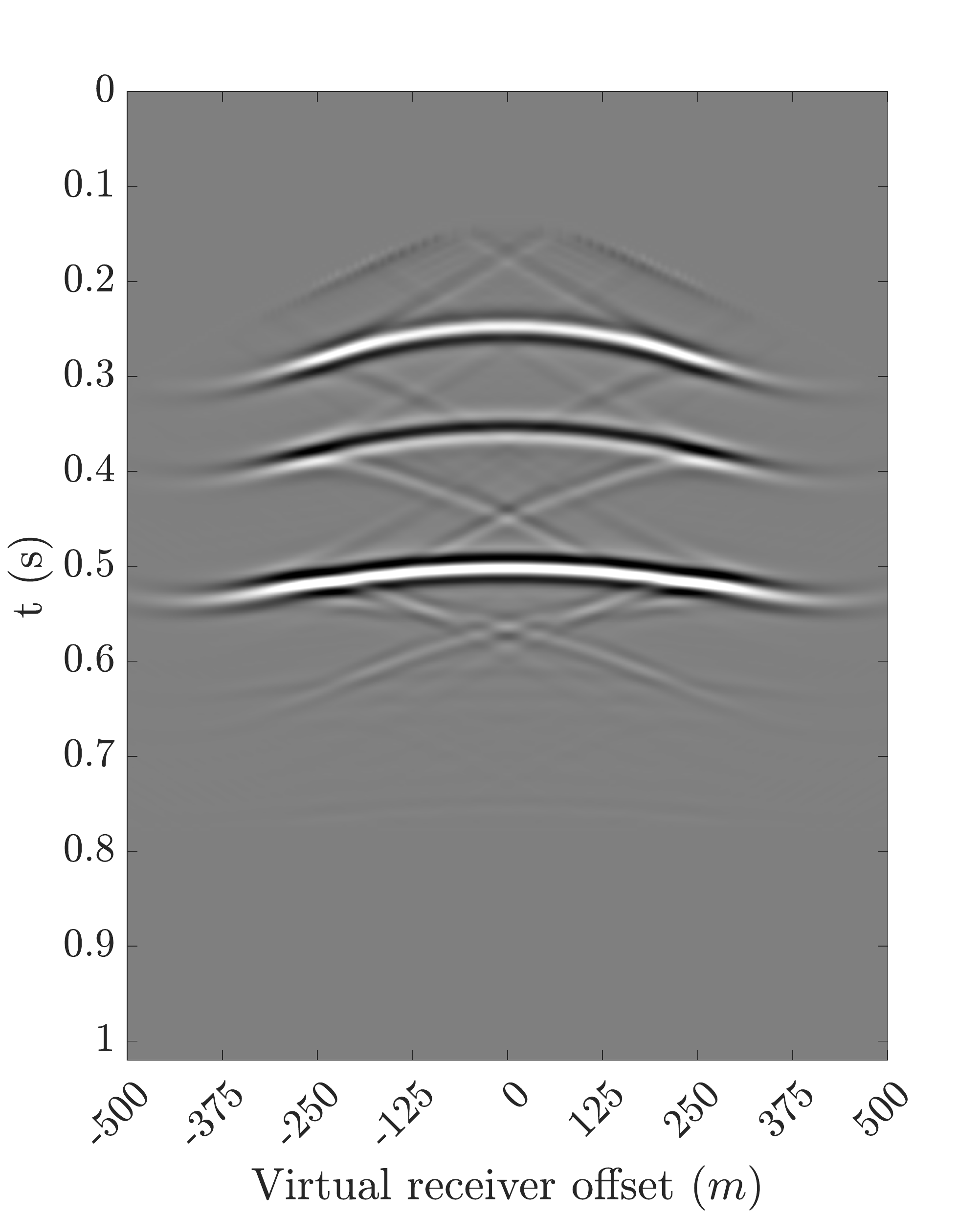}
    \caption{Marchenko redatumed data with virtual receivers at the upper boundary (250 m) and a source located at $\textbf{x}_s=(0,0)$.}
    \label{virobs}
\end{figure}

\begin{figure}
\centering
\begin{subfigure}{0.4\textwidth}
\centering
\includegraphics[width=1\textwidth]{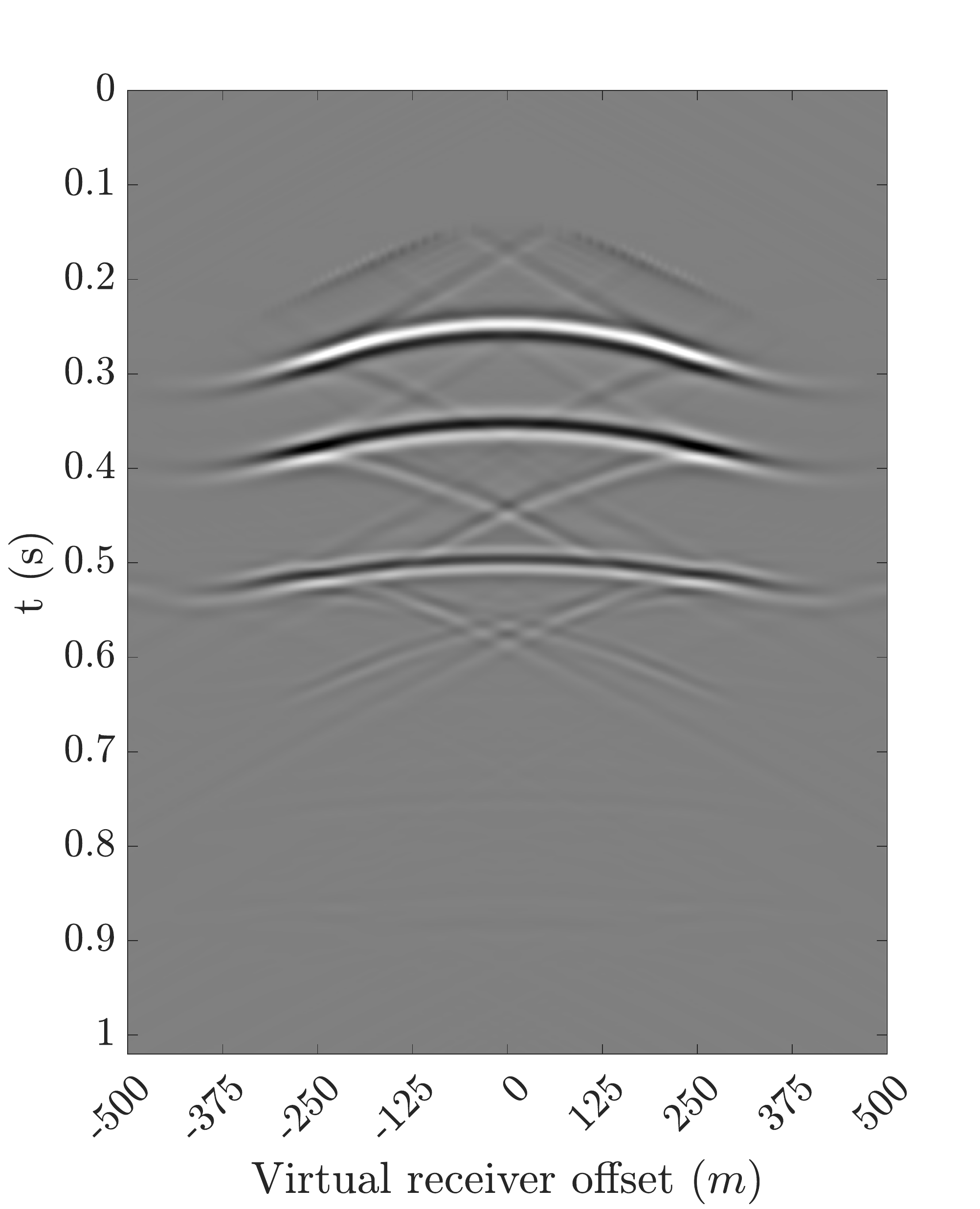}
\caption{}
\label{Marchenko_obs_homogen}
\end{subfigure}
\hfill
\begin{subfigure}{0.4\textwidth}
\centering
\includegraphics[width=1\textwidth]{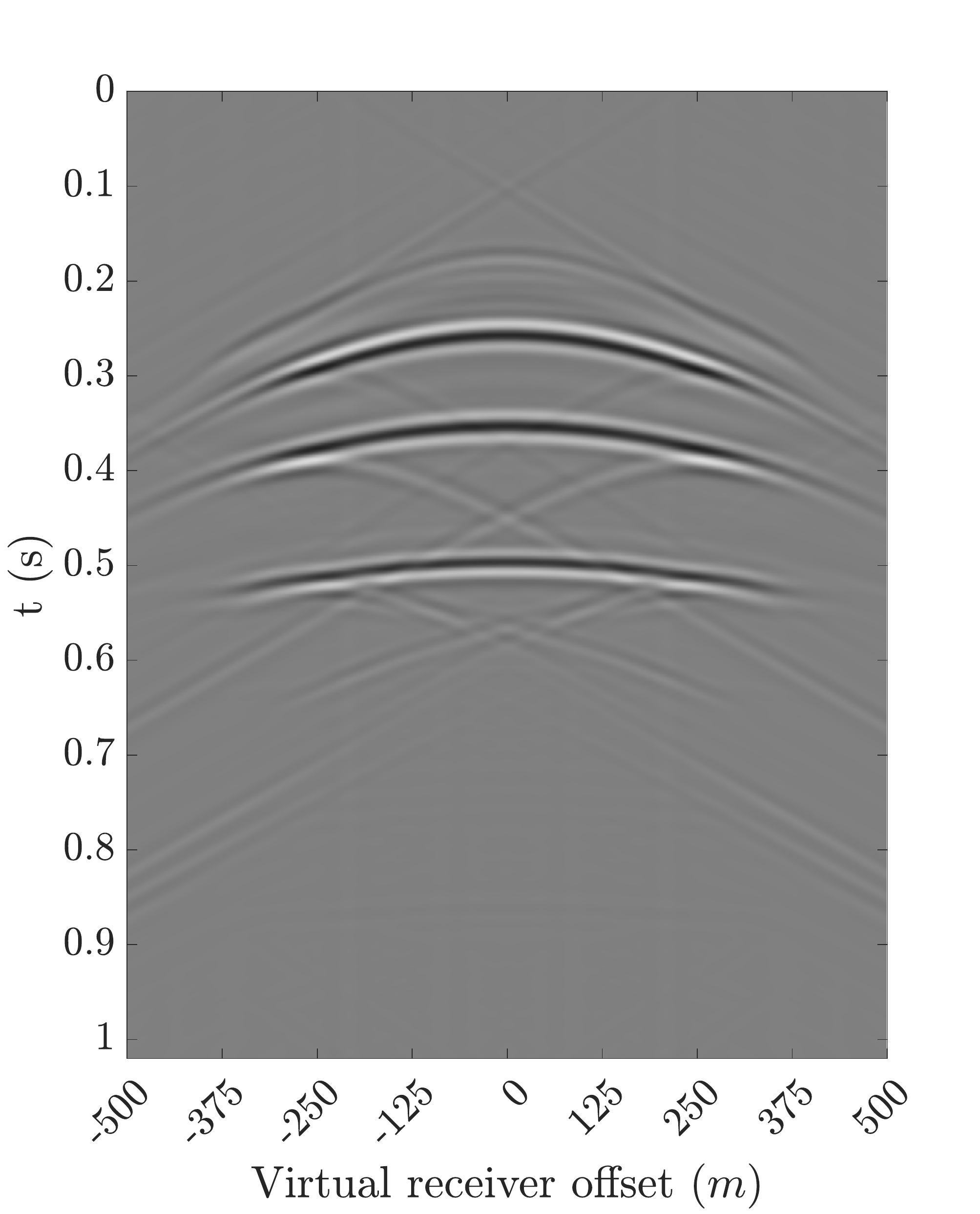}
\caption{}
\label{Marchenko_pred_homogen}
\end{subfigure}
\caption{Homogeneous background velocity case with virtual receivers, data domain: (a) Upgoing component of Marchenko redatumed data at the upper boundary (Equation \ref{PobsTE}), (b) predicted data of target-enclosed LSRTM after 30 iterations at the upper boundary. All wavefields are redatumed to the upper boundary of the target (250 $m$).}
\label{virdata_homogen}
\end{figure}

Fig. \ref{virobs} shows the data obtained by Marchenko redatuming. In Fig. \ref{virdata_homogen} the observed data calculated by Equation \ref{PobsTE} (Fig. \ref{Marchenko_obs_homogen}) and the predicted data after 30 iterations (Fig. \ref{Marchenko_pred_homogen}) are shown. A comparison between Fig. \ref{Marchenko_pred_homogen} and Fig. \ref{pred_enc_homogen} shows that even in the presence of redatuming error, such as limited aperture and lack of certain parts of the wavelength spectrum, our algorithm can predict acceptable data.  

\begin{figure}
\centering
\begin{subfigure}{0.5\textwidth}
\centering
\includegraphics[width=1\textwidth]{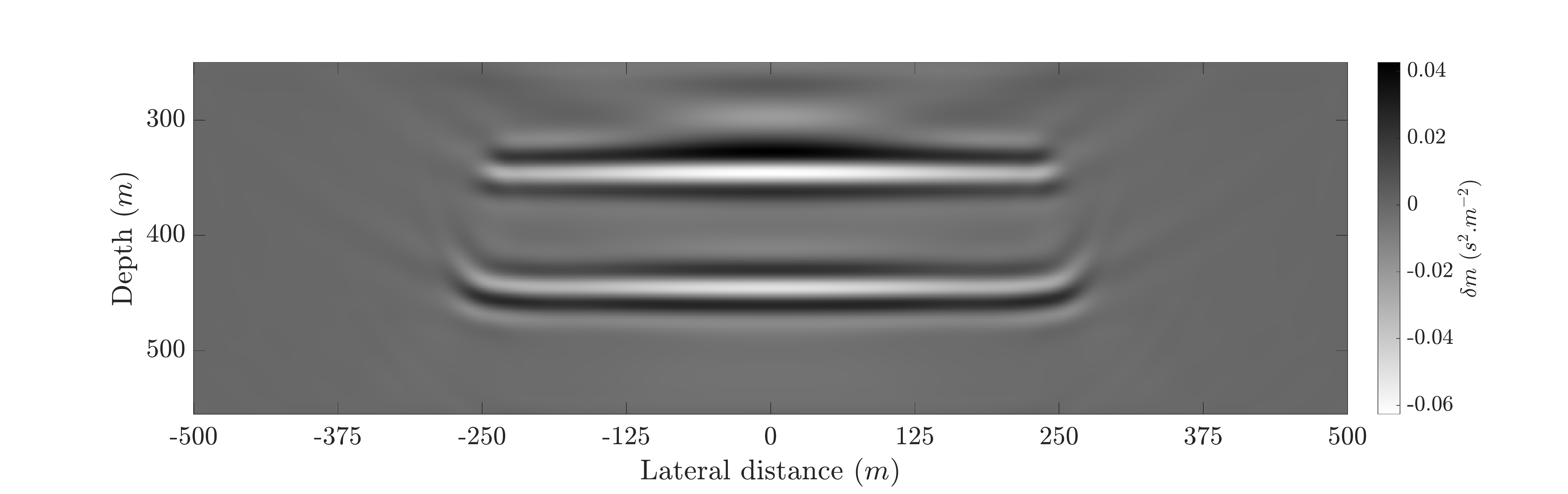}
\caption{}
\label{Marchenko_RTM_homogen}
\end{subfigure}
\hfill
\begin{subfigure}{0.5\textwidth}
\centering
\includegraphics[width=1\textwidth]{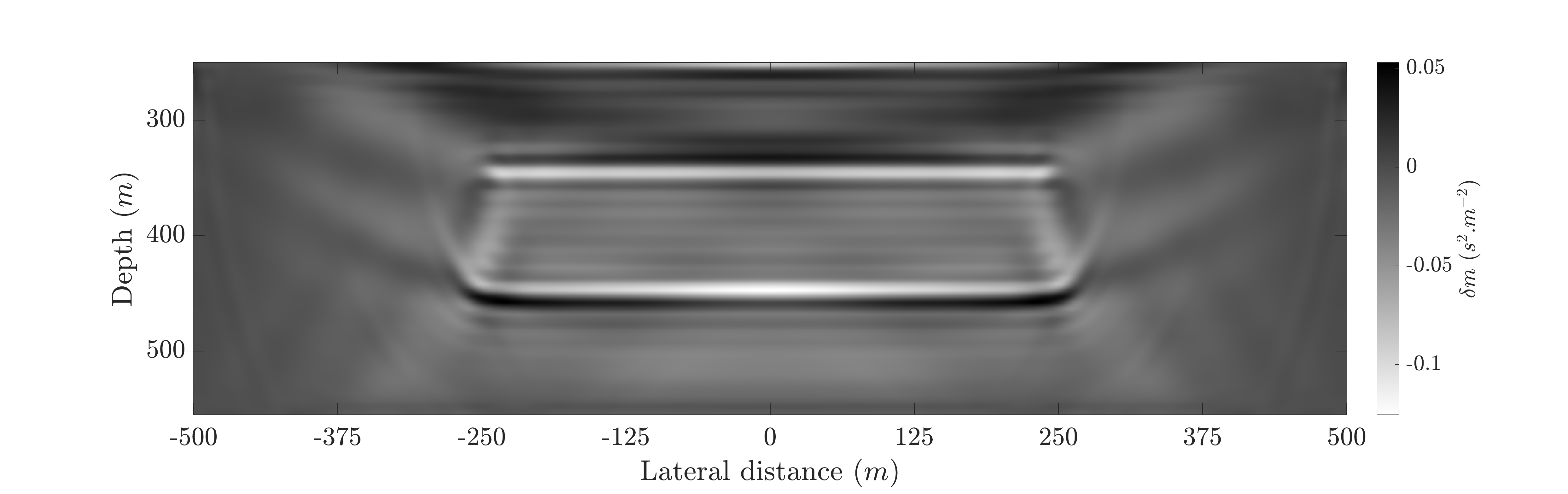}
\caption{}
\label{Marchenko_LSRTM_homogen}
\end{subfigure}
\caption{Homogeneous background velocity case with virtual receivers, image domain: (a) RTM image of the target-enclosed algorithm with Marchenko wavefields, (b) LSRTM image of the target-enclosed algorithm with Marchenko wavefields after 30 iterations.}
\label{virimg_homogen}
\end{figure}

In the image domain, Fig. \ref{virimg_homogen} shows the RTM (Fig. \ref{Marchenko_RTM_homogen}) and LSRTM (Fig. \ref{Marchenko_LSRTM_homogen}) image resulting from Marchenko redatumed data. Comparing Fig. \ref{Marchenko_LSRTM_homogen} with Fig. \ref{LSRTM_Enc_homogen} reveals our target-enclosed algorithm with redatumed data as input can not recover the long wavelength part of the model. This is due to the fact that the direct arrival of the Marchenko-based Green's function is incorrect since it is computed in the background model. Consequently, the forward-scattered waveforms responsible for the long wavelength updates in Figure 7d can no longer be utilized.

\subsubsection{\label{sec:3.2.2} Smooth background velocity}

\begin{figure}
\centering
\begin{subfigure}{0.4\textwidth}
\centering
\includegraphics[width=1\textwidth]{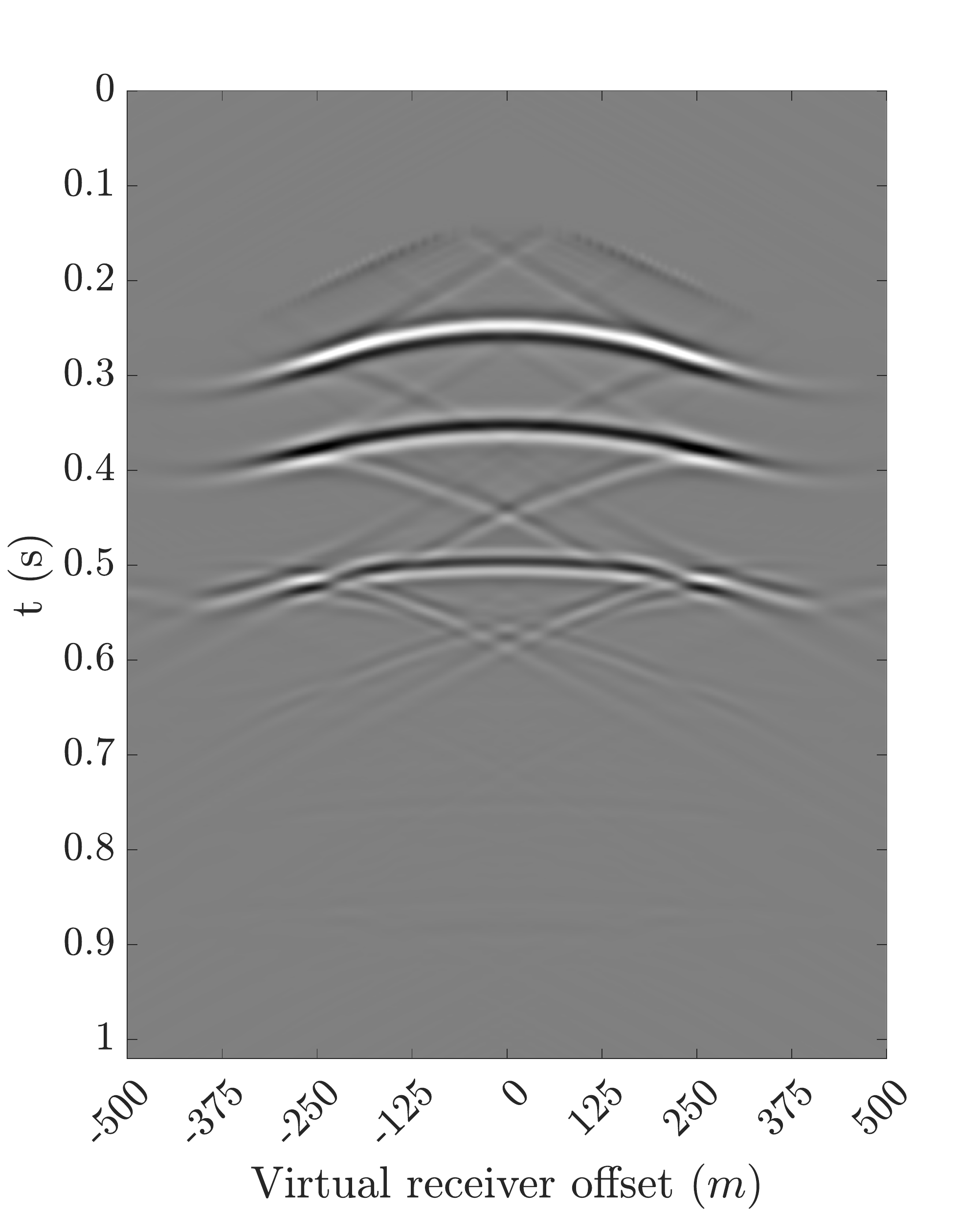}
\caption{}
\label{Marchenko_obs_smooth}
\end{subfigure}
\hfill
\begin{subfigure}{0.4\textwidth}
\centering
\includegraphics[width=1\textwidth]{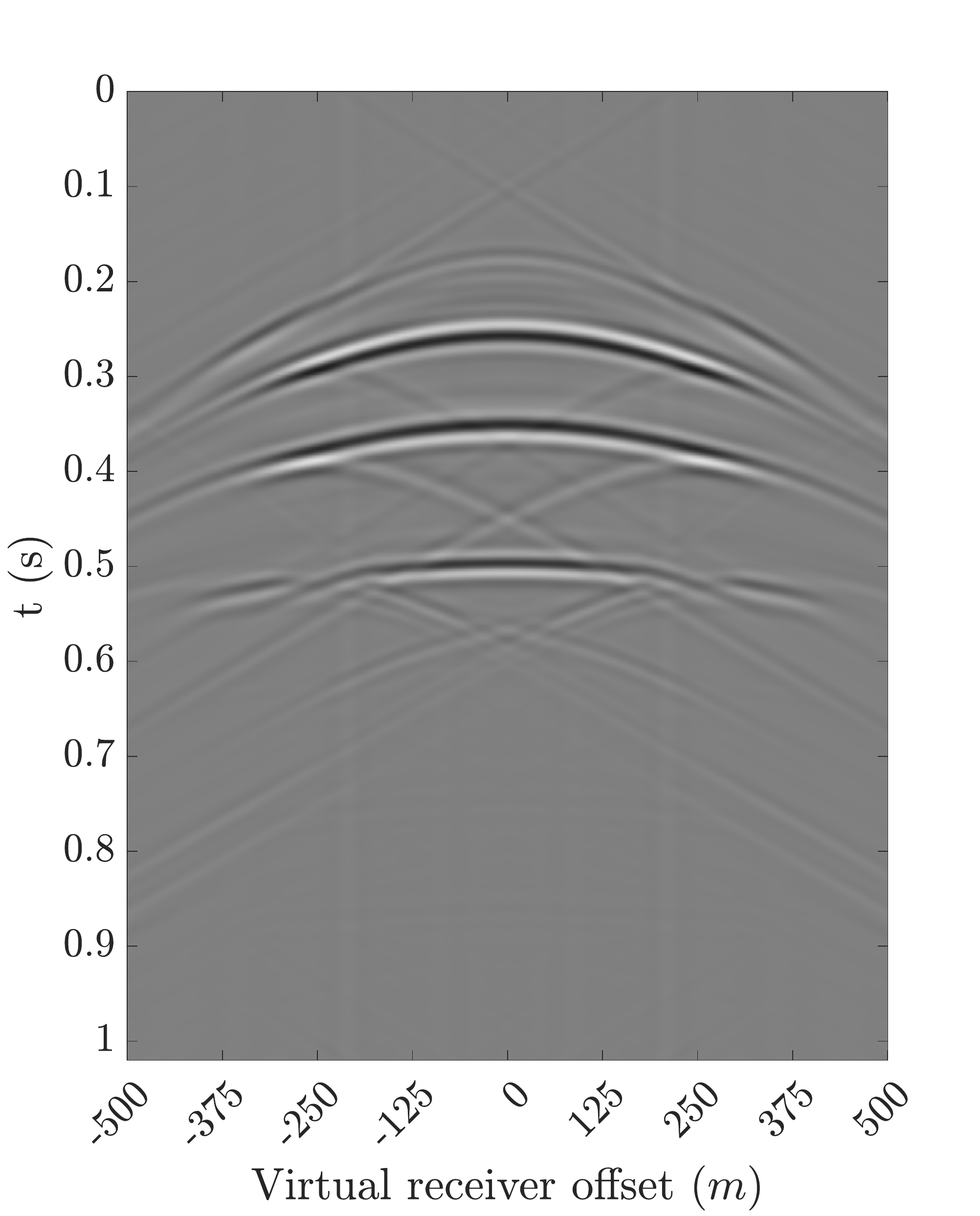}
\caption{}
\label{Marchenko_pred_smooth}
\end{subfigure}
\caption{Smooth background velocity case with virtual receivers, data domain: (a) Upgoing component of Marchenko redatumed data at the upper boundary (Equation \ref{PobsTE}), (b) predicted data of target-enclosed LSRTM after 30 iterations at the upper boundary. All wavefields are redatumed to the upper boundary of the target (250 $m$).}
\label{virdata_smooth}
\end{figure}

In Fig. \ref{virdata_smooth} the observed data calculated by Equation \ref{PobsTE} (Fig. \ref{Marchenko_obs_smooth}) and the predicted data after 30 iterations (Fig. \ref{Marchenko_pred_smooth}) are shown. Similar to the homogeneous case, a comparison between Fig. \ref{virdata_smooth} and Fig. \ref{data_smooth} shows that our algorithm successfully predicts the data.

\begin{figure}
\centering
\begin{subfigure}{0.5\textwidth}
\centering
\includegraphics[width=1\textwidth]{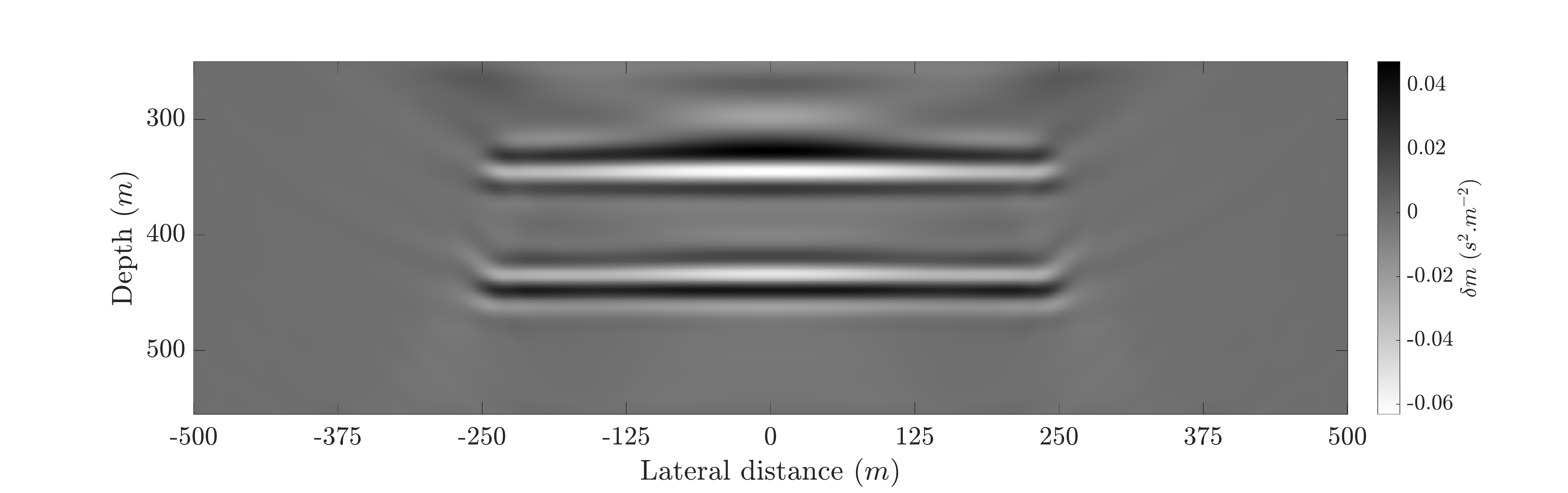}
\caption{}
\label{Marchenko_RTM_smooth}
\end{subfigure}
\hfill
\begin{subfigure}{0.5\textwidth}
\centering
\includegraphics[width=1\textwidth]{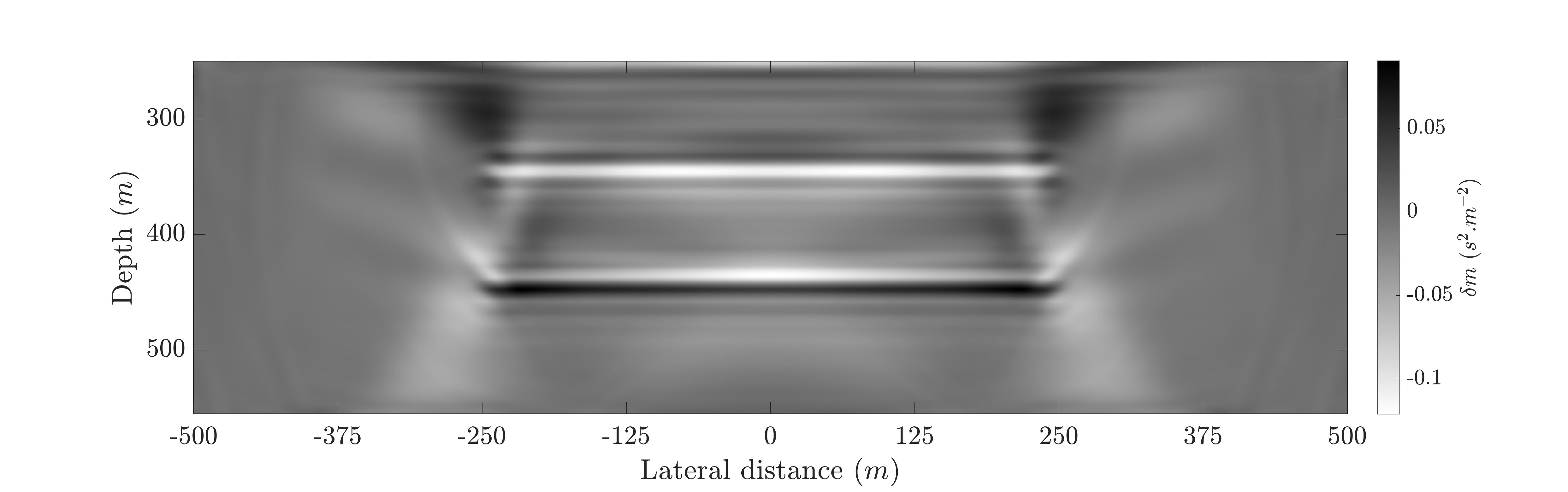}
\caption{}
\label{Marchenko_LSRTM_smooth}
\end{subfigure}
\caption{Smooth background velocity case with virtual receivers, image domain: (a) RTM image of the target-enclosed algorithm with Marchenko wavefields, (b) LSRTM image of the target-enclosed algorithm with Marchenko wavefields after 30 iterations.}
\label{virimg_smooth}
\end{figure}

In the image domain, Fig. \ref{virimg_smooth} shows the RTM (Fig. \ref{Marchenko_RTM_smooth}) and LSRTM (Fig. \ref{Marchenko_LSRTM_smooth}) image resulting from Marchenko redatumed data. Comparing Fig. \ref{Marchenko_LSRTM_smooth} with Fig. \ref{LSRTM_Enc_smooth} shows that redatumed data reveals an acceptable perturbation model. However, the faint recovered vertical interfaces are not presented in Fig. \ref{Marchenko_LSRTM_smooth} since forward-scattered waveforms are not processed correctly in the retrieved Marchenko Green's function at the lower boundary.

\subsubsection{\label{sec:3.2.3} LSRTM results for the entire medium}

\begin{figure}
\centering
\begin{subfigure}{0.5\textwidth}
\centering
\includegraphics[width=1\textwidth]{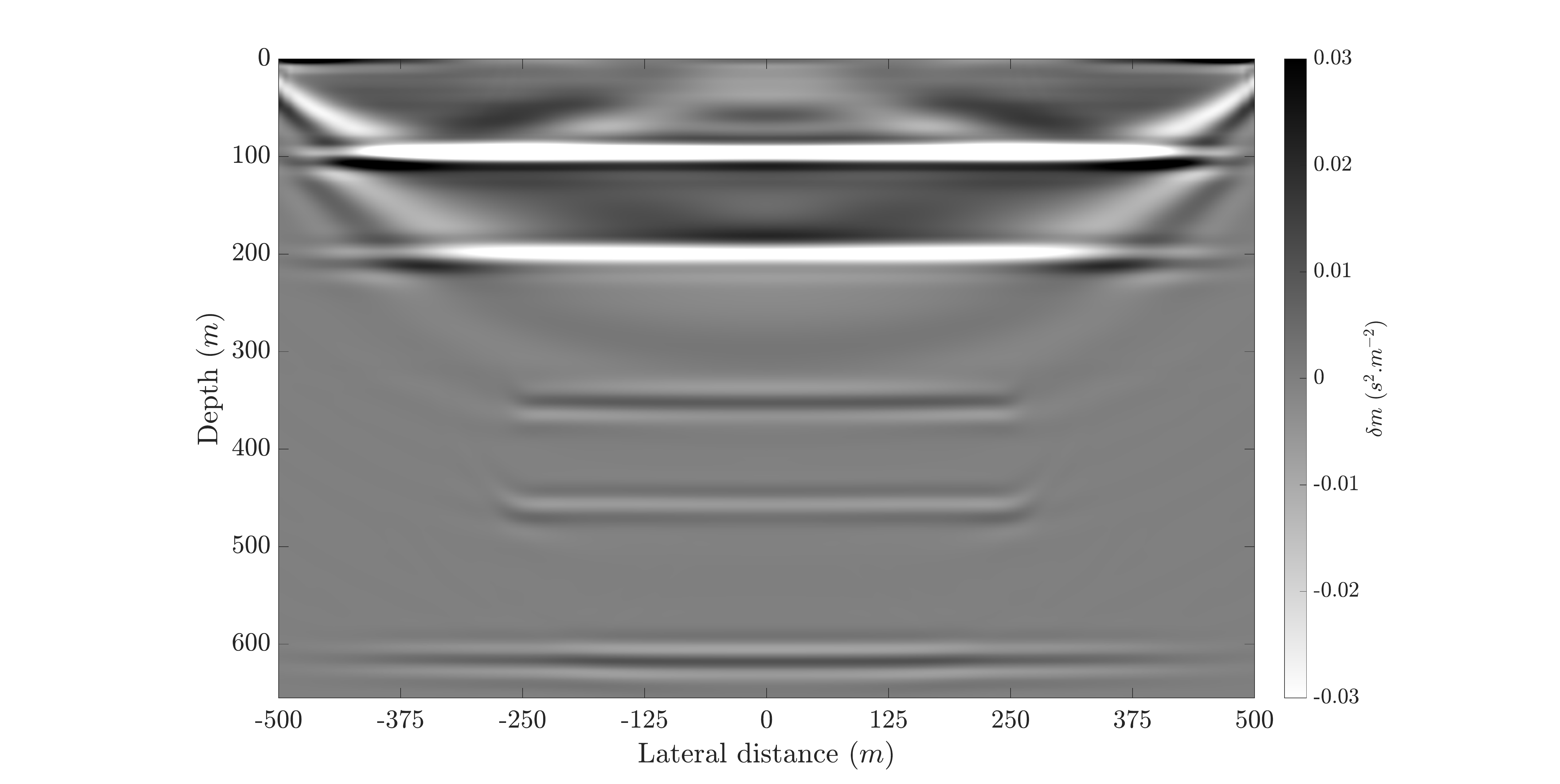}
\caption{}
\label{LSRTM_surf_entire}
\end{subfigure}
\hfill
\begin{subfigure}{0.5\textwidth}
\centering
\includegraphics[width=1\textwidth]{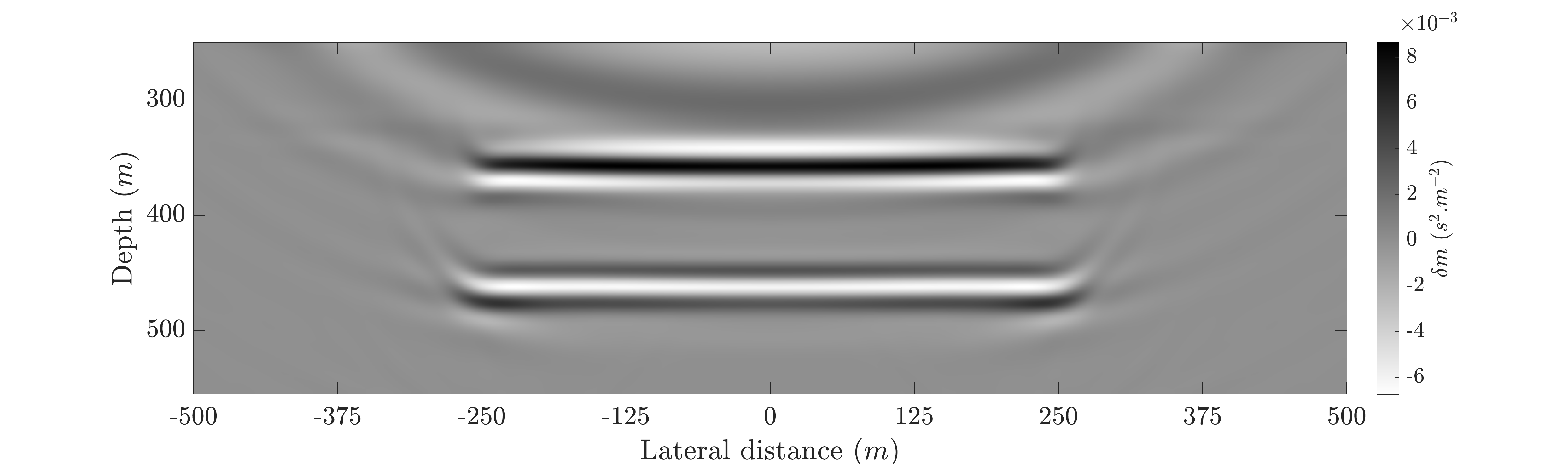}
\caption{}
\label{LSRTM_surf_tar}
\end{subfigure}
\caption{Standard LSRTM with smooth background velocity for entire medium, image domain: (a) LSRTM image of the entire medium, (b) magnified target area of (a).}
\label{LSRTM_surf}
\end{figure}

To make a fair comparison, we include the results of standard LSRTM for the entire medium with the smooth background model in this section. Figure~\ref{LSRTM_surf} shows the LSRTM image after 30 iterations. Figure~\ref{LSRTM_surf_entire} shows the image of the entire medium and Figure~\ref{LSRTM_surf_tar} singled out the target area of it.

\section{Discussion}
In Section \ref{sec:2}, we develop a theory for target-enclosed LSRTM that can limit the computation domain by confining the target between two boundaries. Equations \ref{PobsTE} and \ref{PpredTE} enable us to account for any wavefield entering the target region by including the upper and lower boundaries in the inversion process.

Further, in section \ref{sec:3.1}, several numerical tests are designed to demonstrate the advantage of incorporating the lower boundary in the conventional target-oriented LSRTM. From the data point of view, our double-sided target-enclosed LSRTM, compared to a conventional single-sided target-oriented LSRTM, not only removes the background arrival from the lower boundary to the upper boundary but also can predict the forward scattered field inside the target (Fig. \ref{data_homogen}, Fig. \ref{data_smooth}, and Fig. \ref{virdata_homogen}). Additionally, a comparison between the resulting images of both algorithms with a homogeneous migration velocity (Fig. \ref{img_homogen}), shows that single-sided target-oriented LSRTM can not update the long wavelength part of the model. In contrast, the double-sided target-enclosed LSRTM updates the image according to the Born inversion criteria by integrating the forward scattered wavefield information. In the case of a smooth migration velocity, our algorithm recovers a higher-resolution image and a faint estimation of the vertical sides of the rectangular perturbation (Fig. \ref{img_smooth}).

Moreover, in section \ref{sec:3.2}, we investigate the possibility of using virtual receivers created by Marchenko redatuming in our target-enclosed algorithm. Fig. \ref{virimg_homogen} shows that in the case of a homogeneous background migration velocity, it is hardly possible to update the long wavelengths with virtual receiver data, and only short wavelength parts of the perturbation are recovered. Moreover, for the smooth background case, our algorithm increases the image's resolution by updating the short wavelengths in this case.

Ultimately, we show the standard LSRTM image for the entire medium. Comparing Figure~\ref{LSRTM_surf} with previous cases shows that our method is superior in imaging the target area in any case. Our method adds valuable information by explicitly incorporating transmitted wavefields. Further, with our setup and hardware, the computational time of a single iteration of LSRTM for the entire medium, which has 201 by 131 grid points, is about 45 seconds, and a single iteration of target-enclosed LSRTM, which has 201 by 61 grid points, is about 25 seconds. We do not consider the computational cost of the Marchenko redatuming method here since we only do it once and compared to the total time of LSRTM it is negligible. One disadvantage we can mention is saving the extra redatumed wavefields and focusing functions on the disk. However, relative to the reduction of the memory need by the reduction of the target dimensions, this can be neglected. 

\section{Conclusion}

This paper proposes a target-enclosed seismic imaging algorithm that can account for the wavefields entering the target region from the upper and lower boundaries of the region. The three main advantages of this paper's algorithm are 1) It significantly reduces the computational domain by limiting it to a smaller domain, 2) it removes interactions with the part of the medium above the upper boundary, and 3) it can incorporate the transmission information from the lower boundary to the upper one.

Nevertheless, our algorithm has also some limitations. First, we need access to the lower boundary of the target to deploy receivers at the boundaries of the target. Second, we need a background model of the target that can predict the arrival time from the lower boundary to the upper. It is possible to overcome the first limitation by using virtual seismology methods such as Marchenko redatuming to create virtual receivers around the target region \cite{Broggini,Barrera,Diekman2,Dukalski,Ravasi1,Staring,Wapenaar6,Shoja3} as we showed in a numerical example in section \ref{sec:3.2}. To address the second limitation, a reformulation of the target-enclosed LSRTM is possible to make it compatible with full waveform inversion to update the background velocity model \cite{Shoja}.

The need for high-resolution images is increasing daily, which demands more computational power. This paper's proposed target-enclosed LSRTM can produce less computationally demanding high-resolution images by focusing on a relatively small target of interest and including all the interactions between this region and the outside environment. 


\section*{conflict of interest}
All authors declare that they have no conflicts of interest.

\bibliographystyle{IEEEtran}
\bibliography{IEEEabrv,MyRefs}

\newpage

\section{Biography Section}
 


\begin{IEEEbiography}[{\includegraphics[width=1in,height=1.25in,clip,keepaspectratio]{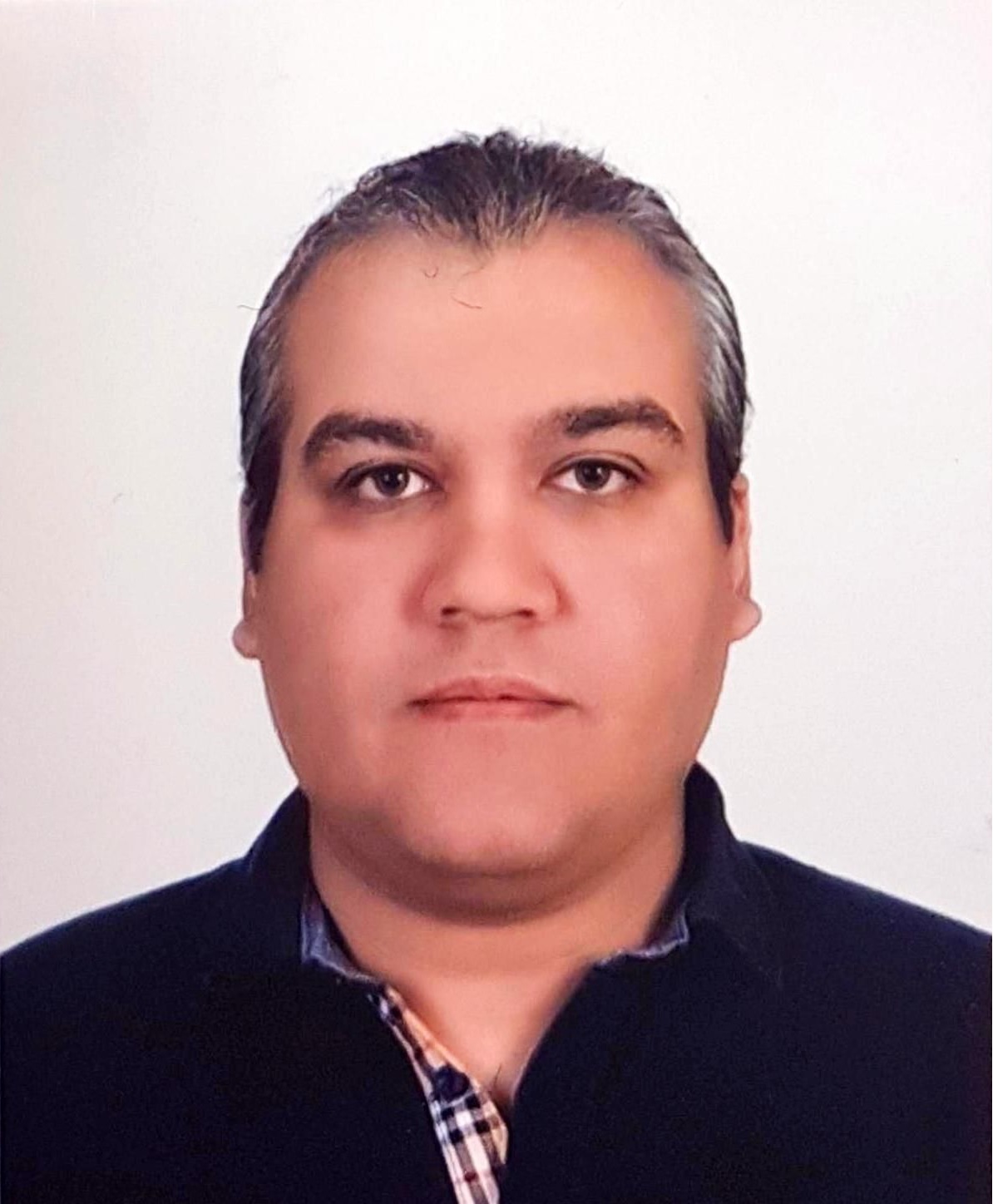}}]{Aydin Shoja}
was born in Kerman, Iran, in 1991. He received the B.Sc. degree in Mining Engineering from the University of Kerman, Kerman, Iran in 2015 and the M.Sc. degree in Geophysics from the University of Tehran, Tehran, Iran in 2018. He is currently a Ph.D. candidate in Applied Geophysics at the Delft University of Technology, Delft, Netherlands. His research has been concerned with wave physics for imaging and inversion purposes.
\end{IEEEbiography}

\begin{IEEEbiography}[{\includegraphics[width=1in,height=1.25in,clip,keepaspectratio]{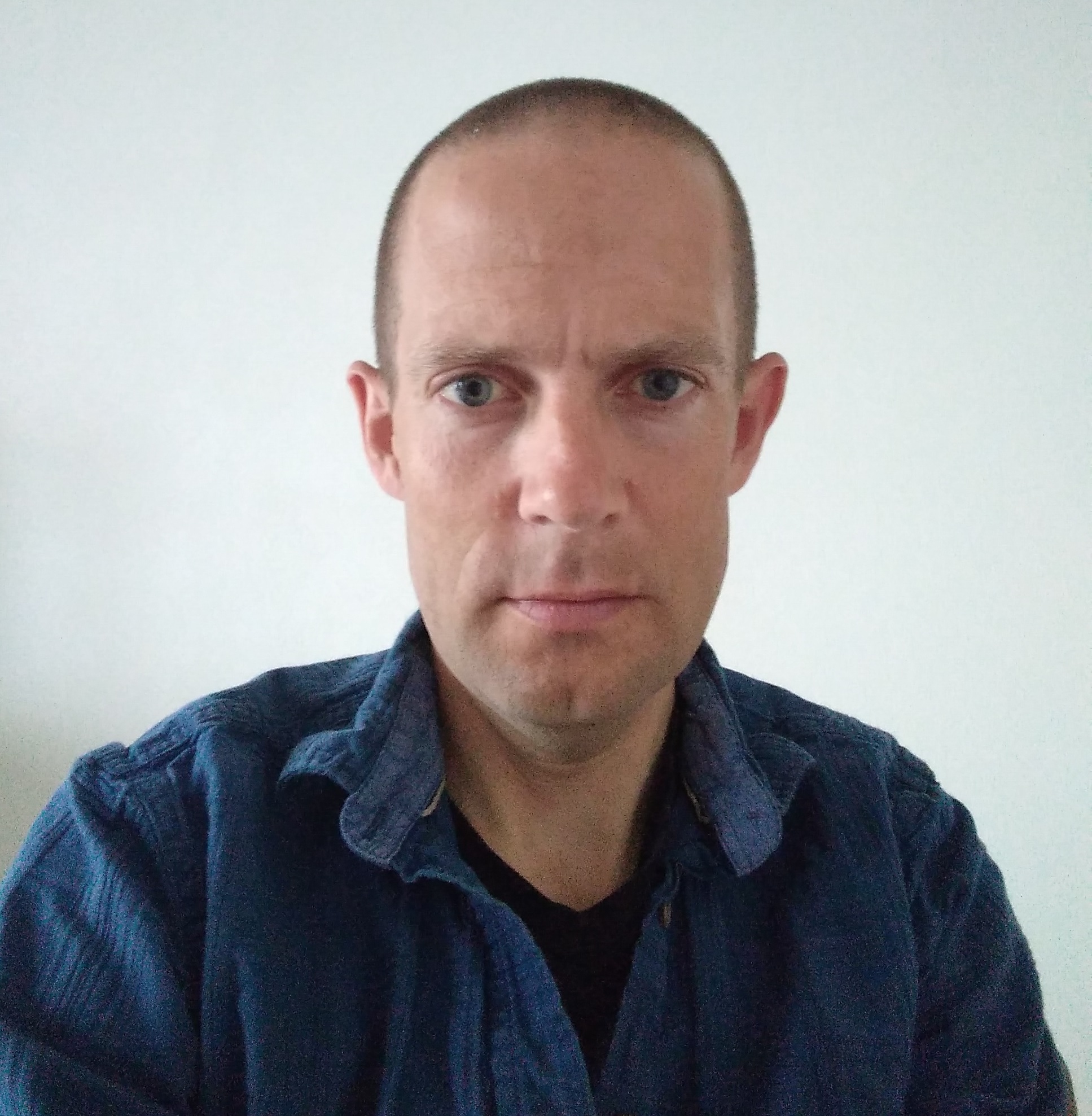}}]{Joost van der Neut}
received the Ph.D. degree (cum laude) in applied geophysics from the Delft University of Technology (TU Delft), Delft, The Netherlands, in 2012. He conducted several postdoctoral studies in seismic and ultrasonic imaging at TU Delft. Dr. van der Neut received a Best Student Presentation Award (SEG, 2009), a Best Presentation Award (SEG, 2010), a J. Clarence Karcher Award (SEG, 2015), and a Best Paper Award (Geophysical Prospecting, 2016). 
\end{IEEEbiography}

\begin{IEEEbiography}[{\includegraphics[width=1in,height=1.25in,clip,keepaspectratio]{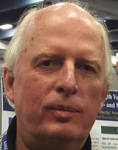}}]{Kees Wapenaar}
received the Ph.D. degree (cum laude) in applied physics from the Delft University of Technology, Delft, The Netherlands, in 1986. From 1986 to 1999, he was a Postdoctoral Research Fellow and an Associate Professor with the Department of Applied Physics, Delft University of Technology. In 1999, he was appointed as Antoni van Leeuwenhoek Professor with the Department of Geoscience and Engineering, where he has been the Head of the Geophysics and Petrophysics Research Group from 2002 until 2022. His research interests include wave theory and its applications in seismic imaging and interferometric methods. Prof. Wapenaar received SEG’s Virgil Kauffman Gold Medal in 2010, EAGE’s Conrad Schlumberger Award in 2013, and a European Research Council (ERC) Advanced Grant from the European Union in 2016. 
\end{IEEEbiography}

\vfill

\end{document}